\documentclass[a4paper,11pt]{article}

\usepackage{bbm, jheppub} 

\usepackage{lineno, xcolor}
\usepackage{csquotes}
\usepackage[T1]{fontenc}
\usepackage{amsthm}
\usepackage{amsmath}

\usepackage{preambles_Saebyeok}

\usepackage{tikz}

\usetikzlibrary{arrows.meta,positioning}

\definecolor{myred}{rgb}{0.75, 0.25, 0}
\definecolor{mygreen}{rgb}{0.25, 0.75, 0}
\definecolor{myblue}{rgb}{0, 0.25, 0.75}

\hypersetup{bookmarksdepth=3, bookmarksnumbered=true, bookmarksopen=false} 

\newcommand{\re}{\mathrm{e}}
\newcommand{\ri}{\mathrm{i}}

\let\originalleft\left
\let\originalright\right
\renewcommand{\left}{\mathopen{}\mathclose\bgroup\originalleft}
\renewcommand{\right}{\aftergroup\egroup\originalright}
\newcommand{\br}[1]{\left( #1 \right)}

\title{Split Heun functions via blown-up surface defects}

\author[a]{Saebyeok Jeong}
\affiliation[a]{Center for Geometry and Physics, Institute for Basic Science (IBS), Pohang 37673,
Korea
}

\author[b,c,d]{and Tommaso Pedroni}
\affiliation[b]{SISSA, Via Bonomea 265, 34136 Trieste, Italy}
\affiliation[c]{INFN, Sezione di Trieste, Trieste, Italy}
\affiliation[d]{Institute for Geometry and Physics, IGAP, via Beirut 2, 34136 Trieste, Italy}

\emailAdd{saebyeok.jeong@gmail.com}
\emailAdd{tpedroni@sissa.it}

\abstract{
We study resonant solutions of the Heun equation and its confluent limits that arise in the Nekrasov--Shatashvili (NS) limit of four-dimensional $\mathcal{N}=2$ $\mathrm{SU}(2)$ gauge theories with fundamental hypermultiplets. At the resonant loci $2a/\hbar\in\mathbb{Z}$ in the Coulomb branch parameter $a$, the Floquet multipliers coalesce and the instanton expansions of the bulk and surface defect NS functions develop poles of increasing order. We derive blow-up equations involving exclusively NS functions and use them to resum these singular expansions. The resulting resummed bulk and surface defect NS functions reveal the analytic structure of the gauge-theoretic solutions near the resonant loci, including the branch structure of the accessory parameter and of the Floquet solutions that is obscured by the term-by-term instanton expansion. At resonance, the resummed accessory parameters and suitably normalized defect wavefunctions admit finite limits that describe periodic or antiperiodic solutions at the edges of spectral gaps and allow us to construct their logarithmic companions. We then identify distinct nested mass loci governing gap closure and semisimple resonant monodromy. On the larger locus the band-edge accessory parameters coalesce, while on the smaller locus two independent resonant (anti)periodic Floquet solutions survive. We develop the general resummation procedure for $N_f=(n_0,n_1)$ theories with $n_i\leq 2$ $(i=0,1)$, and demonstrate it explicitly for the $N_f=(1,1)$ theory.}

\begin{document}

\phantomsection 
\pdfbookmark[1]{Abstract}{Abstract} 

\maketitle

\flushbottom

\newpage
\section{Introduction}

The four-dimensional $\CalN=2$ supersymmetric gauge theories of class $\CalS$ \cite{Gaiotto:2009we} provide a powerful framework for studying geometric structures on the associated Riemann surface. In the Nekrasov--Shatashvili (NS) limit of the $\Omega_{\ve_1,\ve_2}$-background \cite{Nekrasov:2002qd}, $\varepsilon_1 = \hbar$ and $\ve_2\to0$ \cite{Nekrasov:2009rc}, the Seiberg--Witten curve \cite{Seiberg:1994rs} of the gauge theory is promoted to a differential operator \cite{Nekrasov:2015wsu, Nekrasov:2017gzb}, an oper \cite{1885368}, on the Riemann surface. The locus of such opers forms a holomorphic Lagrangian submanifold of the moduli space of parabolic local systems. In the NS limit, the effective twisted superpotential $\mathcal{W}_0$ provides a generating function for this Lagrangian submanifold in a certain Darboux coordinate system, with the Coulomb parameters supplying one half of the Darboux coordinates \cite{Nekrasov:2011bc,Jeong:2018qpc}. For the $\mathrm{SU}(2)$ gauge theory with $N_f=4$ fundamental hypermultiplets, the relevant Riemann surface is the sphere $\BP^1$ with four marked points $S=\{0,\qe,1,\infty\}$, and the corresponding $\mathrm{PGL}(2)$-oper is precisely the celebrated Heun differential operator.

The correspondence is most directly realized over the generic locus of the Coulomb moduli space \cite{Jeong:2018qpc}. A weakly coupled class $\CalS$ theory selects a pants decomposition of the four-punctured sphere, obtained by gluing two three-punctured spheres through a tube. The eigenvalues of the oper monodromy along the curve separating the two pairs of pants, which we call the $A$-cycle, are determined by the Coulomb parameter $a$. Accordingly, the accessory parameter of the Heun operator is determined by $a$ through the effective twisted superpotential. A basis of Floquet solutions to the Heun equation can likewise be constructed from the expectation values of regular half-BPS orbifold surface defects in the same limit \cite{Kanno:2011fw,Jeong:2018qpc}. At the resonant loci
\[
    \frac{2a}{\hbar}=n\in\BZ,
\]
however, the two $A$-cycle Floquet multipliers $\re^{\pm2\pi\ri a/\hbar}$ coalesce to $(-1)^n$. This is a global $A$-cycle resonance, rather than a collision of local exponents at one of the punctures. For generic values of the fundamental hypermultiplets' masses, the resulting monodromy is not semisimple and the second independent solution is logarithmic.

The obstruction to applying the usual gauge-theoretic construction at a nonzero resonance is already visible in the instanton expansion. The coefficients of the bulk NS functions $\mathcal{W}_0$ and $\mathcal{W}_1$, as well as those of the defect NS functions $\hat{\mathcal{W}}_\beta$, see \eqref{eq: NS expansion} for their definition, develop poles at $2a=\pm j\hbar$, $j\in\mathbb{Z}_{>0}$, whose order grows with the instanton number \cite{Beccaria:2016wop,Jeong:2017pai,Gorsky:2017ndg,Bonelli:2025bmt}. The resonance $a=0$ is exceptional, since the instanton functions considered here remain regular there and the resonant limit can be taken term by term. Consequently, neither the accessory parameter obtained from the quantum Matone relation \cite{Matone:1995rx,Flume:2004rp} nor the defect wavefunctions possess a term-by-term resonant limit: the instanton expansion and the resonant limit cannot be interchanged.

For the pure $\mathrm{SU}(2)$ theory, complementary aspects of this problem for the Mathieu operator were previously resolved in \cite{Jeong:2017pai,Gorsky:2017ndg}. In \cite{Jeong:2017pai}, suitable combinations of orbifold surface defect partition functions were shown to split at the resonant locus, yielding both the split band-edge eigenvalues and the corresponding Mathieu eigenfunctions (and their higher-rank generalizations). In \cite{Gorsky:2017ndg}, building on \cite{Beccaria:2016wop}, a resummation ansatz was formulated for the singular instanton expansion of the bulk NS twisted superpotential, reorganizing the naive poles into branch cuts and recovering the band-edge eigenvalues through the quantum Matone relation. 

Finally, a more systematic resummation procedure based on the blow-up equations \cite{Nakajima:2003pg} satisfied by bulk Nekrasov functions was developed in \cite{Bonelli:2025bmt}. This framework naturally encompasses both the $\mathcal{N}=2^*$ $\mathrm{SU}(2)$ gauge theory, also previously studied in \cite{Beccaria:2016wop,Gorsky:2017ndg}, and $\mathrm{SU}(2)$ gauge theories with fundamental hypermultiplets. These developments do not yet provide the simultaneous resummation of bulk and defect NS functions in the presence of fundamental hypermultiplets that is needed to construct the corresponding resonant Floquet solutions. Moreover, the standard collision construction of their logarithmic companions can only be applied after the singular instanton expansions have been resummed.

The main purpose of this work is to construct the resonant accessory parameters and the corresponding split eigenfunctions associated with the Heun equation and its confluent limits. Our principal tool is a new class of blow-up equations involving exclusively bulk and defect NS functions. Starting from the blow-up equations for orbifold surface defect partition functions \cite{Jeong:2020uxz}, which generalize their bulk counterparts \cite{Nakajima:2003pg,Nekrasov:2020qcq}, we take a suitable limit of the equations associated with two distinct values of the flux along the exceptional divisor, following the approach developed in \cite{Bonelli:2025bmt}. Comparing the two equations and eliminating their common part, we are led to a new class of blow-up equations involving exclusively the NS functions $\mathcal{W}_0$, $\mathcal{W}_1$, and $\hat{\mathcal{W}}_\beta$. We first derive these equations for the $N_f=(2,2)$ theory, see \eqref{eq: blowup combined (2,2)}, where the notation records the flavor coloring associated with the orbifold, and then obtain their $N_f=(n_0,n_1)$ counterparts by decoupling hypermultiplets, see \eqref{eq: blowup combined}. The latter govern the corresponding confluent degenerations of the Heun equation.

To solve these equations near resonance, we set $\qe=\mathfrak{t}^2$ and reorganize all pole contributions associated with the $j$-th resonance into functions of the correlated variables
\[
   x_j^{\pm} = \frac{\mathfrak{t}^j}{j\pm2a/\hbar}.
\]
This resums entire towers of Coulomb-branch poles across different instanton orders. Given the resummation functions for the bulk NS functions, the new blow-up equations recursively determine their defect counterparts. 

Based on the computation of the first two $j=1$ resummation functions for $N_f=(2,2)$, the first three $j=1$ and the first two $j=2$ functions for $N_f=(1,1)$, and a slightly larger set for $N_f=(0,0)$, we conjecture closed-form expressions for the leading defect resummation functions at arbitrary resonance, see \eqref{eq: resumm leading h} and \eqref{eq: resumm leading ht}. We have tested these conjectured expressions against the instanton expansion of the defect NS functions through order $\mathfrak{t}^{10}$.

The increasing towers of poles are thereby replaced by controlled square-root and logarithmic branch behavior. Although the resummed NS functions are not holomorphic at the resonant points, the resummed accessory parameter and suitably normalized defect wavefunctions possess finite one-sided limits.

Away from resonance, the resummed defect partition functions furnish a basis of Floquet solutions. At a nonzero resonant value $2a/\hbar=n$, the resummed accessory parameter generally admits two lateral limits,
\[
    u_{\sigma}^{(n)}
    :=
    \lim_{2a/\hbar\to n^{\sigma}}
    u_{\mathrm{res}}(a),
    \qquad
    \sigma\in\{+,-\}.
\]
The corresponding limits of the suitably normalized resummed defect functions give periodic or antiperiodic solutions $\phi_{\sigma}^{(n)}$. At generic masses, $u_+^{(n)}\neq u_-^{(n)}$, so these solutions belong to two different resonant Heun operators. On an appropriate real slice, the two accessory parameters are naturally interpreted as the two edges of the spectral gap associated with the resonance.

For each of these resonant Heun operators, the $A$-cycle monodromy is generically non-semisimple, and the second independent solution is logarithmic. We show how to obtain this logarithmic companion both by reduction of order and by taking the collision limit of the two resummed Floquet solutions. The coefficient $K_{\sigma}^{(n)}$ of the logarithmic term gives a local representative of the nilpotent part of the resonant $A$-cycle monodromy,
\[
M_A
=
(-1)^n
\begin{pmatrix}
1 & 0 \\
2\pi i\,K_{\sigma}^{(n)} & 1
\end{pmatrix}.
\]
Although its value depends on the normalization of $\phi_{\sigma}^{(n)}$, its vanishing is invariant under regular changes of normalization and characterizes semisimplicity of the corresponding branch.

We further conjecture that, on each lateral branch, the resummed dual quantum period satisfies 
\[
\lim_{2a/\hbar\to n^{\sigma}}
    \exp \br{a_D} =1.
\]
This in turn implies
\[
    \lim_{2a/\hbar\to n^{\sigma}}
    \partial_a u_{\mathrm{res}}=0,
\]
and, under the appropriate normalization and regularity assumptions, allows the logarithmic companion to be represented by the resonant limit of the $a$-derivative of a single resummed Floquet solution, see the end of Section \ref{sec:logsol}.

We have tested this conjecture using the first two $j=1$ resummation functions and the first $j=2$ resummation function for $N_f=(2,2)$, the first three $j=1$, the first two $j=2$, and the first two $j=3$ resummation functions for $N_f=(1,1)$, and a slightly larger set for $N_f=(0,0)$.

Finally, the pattern of pole cancellation in the instanton expansions of both the accessory parameter and the Floquet solutions identifies special loci in the space of mass parameters at which the corresponding spectral gaps close. In units where $\hbar=1$, these loci are determined by the zeros of
\[
    \mathfrak p_N(m)
    =
    \prod_{k=1-\frac N2}^{\frac N2}(m+k),
    \qquad N=|n|.
\]

Complete cancellation of the poles in the bulk instanton expansions requires two distinct mass parameters to be specialized to zeros of $\mathfrak p_N$. This defines the locus $\mathcal D_{\mathrm{bulk}}^{(N)}$, on which the two band-edge accessory parameters coalesce,
\[
u_+^{(n)}=u_-^{(n)}, \qquad N=|n|,
\]
and hence the $N$-th spectral gap closes. Regularity of both defect solutions imposes the stronger condition that at least one mass of each flavor color be specialized. For the $N_f=(2,2)$ theory, this means choosing one mass from $(m_1,m_2)$ and one from $(m_3,m_4)$. The resulting locus $\mathcal D_{\mathrm{defect}}^{(N)}$ is contained in
$\mathcal D_{\mathrm{bulk}}^{(N)}$. On this smaller locus, the two appropriately normalized resonant defect wavefunctions remain linearly independent and solve the same Heun equation, so that
\[
M_A=(-1)^N I.
\]
By contrast, on $\mathcal D_{\mathrm{bulk}}^{(N)} \setminus\mathcal D_{\mathrm{defect}}^{(N)}$, the gap closes but the two defect wavefunctions yield the same resonant (anti)periodic Floquet solution after renormalization, the second independent solution being logarithmic. Thus gap closure does not by itself imply semisimplicity.

Within the small-$\mathfrak t$ analysis, the semisimple locus is identified with $\mathcal D_{\mathrm{defect}}^{(N)}$. Since these loci are nested separately for odd and even $N$, the corresponding mass specializations close infinitely many gaps of one parity without, in general, defining finite-gap potentials.

The paper is organized as follows. In Section~\ref{sec:orbifold}, we review regular orbifold surface defects and state the differential equations and the blow-up equations they satisfy. We also present the corresponding hypermultiplet decoupling limits. In Section~\ref{sec:resummation}, we develop the resummation procedure for Floquet solutions, construct the resonant and logarithmic solutions, and analyze the special mass loci associated with semisimple monodromy. We conclude the analysis with an explicit illustration in the $N_f=(1,1)$ theory. The Appendices \ref{app: special functions} and \ref{app: example} contain a brief review of the relevant bulk Nekrasov functions and some computational details for the $N_f = (1,1)$ theory.

\section{Blow-up equations for orbifold surface defects}
\label{sec:orbifold}

In this section, we first review the orbifold construction of surface-defect partition functions in four-dimensional $\mathcal{N}=2$ gauge theories, focusing in particular on the $\mathrm{SU}(2)$ theory with $N_f=4$, which we will denote by $N_f=(2,2)$ for reasons that will become clear below. We then discuss the differential equations satisfied by the corresponding defect partition functions, together with the associated blow-up equations. In particular, we consider a special limit of the latter in which only Nekrasov--Shatashvili functions survive. Finally, we discuss the decoupling limits of the $N_f=(2,2)$ gauge theory at the level of both the partition functions and the differential/blow-up equations.

\subsection{Orbifold surface defects and solutions of the Heun equation}
\label{sec:orbifoldconstr}
There are several ways to define BPS surface defects in $\mathcal{N}=2$ gauge theories \cite{Gaiotto:2009fs}. One approach is to prescribe singular boundary conditions for the fields along the surface, which can be modeled using an orbifold construction \cite{Kanno:2011fw}. Consider an $\mathcal{N}=2$ gauge theory on the orbifold $\mathbb{C} \times (\mathbb{C} /\mathbb{Z}_p)$, defined by the following $\mathbb{Z}_p$ action on $\mathbb{C}^2$:
\begin{equation}
    \zeta : \left( z_1, z_2\right) \mapsto \left( z_1, \zeta z_2\right), \quad \zeta = \exp\left( \frac{2\pi \ri}{p}\right) \in \mathbb{Z}_p,
\end{equation}
with the orbifold singularity located along the $z_1$-plane. The orbifold can be mapped to $\mathbb{C}^2$ via the change of variables
\begin{equation}
    \left( z_1, z_2\right) \mapsto (\tilde{z}_1=z_1,\tilde{z}_2=z_2^p),
\end{equation}
in terms of which the field configurations are allowed to develop singularities along the surface $\tilde{z}_2=0$, thereby engineering a BPS surface defect. 

To fully characterize the surface defect, one must specify a choice of coloring function
\begin{equation}
    c:[N]\longrightarrow [p],
\end{equation}
where $[N]:=\{0,1,\ldots,N-1\}$ and $[p]:=\{0,1,\ldots,p-1\}$. This specifies the lift of the $\mathbb Z_p$-action to the gauge framing space and hence determines the Levi subgroup of the gauge symmetry preserved by the defect.

In the presence of fundamental hypermultiplets, one must also specify a lift of the orbifold action to their flavor space. This is encoded by a flavor-coloring function
\begin{equation}
    c_{\mathrm f}:[N_f]\longrightarrow[p].
\end{equation}
Equivalently, if $M \simeq\mathbb C^{N_f}$ denotes the flavor space, the coloring determines a $\mathbb Z_p$-grading
\begin{equation}
    M=\bigoplus_{\omega\in\mathbb Z_p} M_\omega,
    \qquad
    n_\omega:=\dim  M_\omega
    =\bigl|c_{\mathrm f}^{-1}(\omega)\bigr|.
\end{equation}
The masses assigned color $\omega$ are the equivariant weights of the flavor torus acting on $ M_\omega$. We will use the notation,
\begin{equation}
    N_f=(n_0,n_1,\ldots,n_{p-1}),
\end{equation}
for the resulting colored matter content, so that the total number of fundamental hypermultiplets is $\sum_{\omega}n_\omega$. In the $\mathbb Z_2$ case relevant below, this becomes $N_f=(n_0,n_1)$, where $n_0$ and $n_1$ count the fundamental hypermultiplets of colors $0$ and $1$, respectively. For the parent theory with four fundamental hypermultiplets, we choose $(m_3,m_4)$ to have color $0$ and $(m_1,m_2)$ to have color $1$, and hence denote the theory by $N_f=(2,2)$.

The Nekrasov instanton partition function of the $\mathcal{N}=2$ gauge theory in the presence of such a defect is obtained by implementing the $\mathbb{Z}_p$ orbifold at the level of the ADHM construction. In this formulation, the ADHM data are decomposed according to the $\mathbb{Z}_p$-weights, yielding a chain-saw quiver with $p$ nodes \cite{Kanno:2011fw}. 

For the purpose of studying the Heun oper and its degenerations, we restrict our attention to $\mathrm{SU}(2)$ gauge theories with fundamental hypermultiplets. We also take the orbifold surface defects to be \textit{regular}, which are characterized by a one-to-one coloring function (in particular, $p=2$). In this case, there are only two possible choices, namely $c=\mathrm{id}$ and $c=(01)$, the elementary transposition. 

\begin{figure}
    \centering
    \begin{tikzpicture}[
  >={Stealth[length=2mm]},
  vtx/.style={circle, draw, minimum size=8mm, inner sep=0pt},
  wtx/.style={draw, minimum width=8mm, minimum height=8mm, inner sep=0pt},
  arr/.style={->, line width=0.5pt},
  lab/.style={font=\small, fill=white, inner sep=1pt}
]

  \node[vtx] (V1) {$V_0$};
  \node[vtx, right=3.0cm of V1] (V2) {$V_1$};

  \node[wtx, below=1.5cm of V1] (W1) {$W_0$};
  \node[wtx, below=1.5cm of V2] (W2) {$W_1$};

  \draw[arr] (V1) to[out=130,in=50,looseness=8]
    node[lab, above] {$A_0$} (V1);
  \draw[arr] (V2) to[out=130,in=50,looseness=8]
    node[lab, above] {$A_1$} (V2);

 \draw[arr] ([yshift=1.3mm]V1.east) --
    node[lab, above] {$B_0$}
    ([yshift=1.3mm]V2.west);

  \draw[arr] ([yshift=-1.3mm]V2.west) --
    node[lab, below] {$B_1$}
    ([yshift=-1.3mm]V1.east);

  \draw[arr] (W1) -- node[lab, left] {$P_0$} (V1);
  \draw[arr] (W2) -- node[lab, right] {$P_1$} (V2);

  \draw[arr] (V1) -- node[lab, below, sloped, pos=0.28] {$Q_0$} (W2);
  \draw[arr] (V2) -- node[lab, below, sloped, pos=0.28] {$Q_1$} (W1);
\end{tikzpicture}
    \caption{$\BZ_2$ chain-saw quiver}
    \label{fig:csquiver}
\end{figure}

The chain-saw quiver is drawn in Figure \ref{fig:csquiver} and in the present $\BZ_2$ case, all indices are understood modulo $2$. Let $V_\omega$ and $W_\omega$ be vector spaces of dimensions
\begin{align} \nonumber
    \dim V_\omega = k_\omega,
    \quad 
    \dim W_\omega = 1,
    \qquad 
    \omega \in \BZ_2 .
\end{align}
A representation of the $\BZ_2$ chain-saw quiver consists of linear maps between these vector spaces:
\[
    A_\omega \in \text{End}(V_\omega), 
    \quad
    B_\omega \in \text{Hom}(V_\omega,V_{\omega+1}), \quad
    P_\omega \in \text{Hom}(W_\omega,V_\omega),
    \quad
    Q_\omega \in \text{Hom}(V_\omega,W_{\omega+1}).
\]
The chain-saw moment map equations are
\[
   \m_\o:= A_{\omega+1}B_\omega - B_\omega A_\omega + P_{\omega+1}Q_\omega = 0,
    \qquad \omega \in \BZ_2,
\]
or, explicitly,
\begin{equation*}
\begin{aligned}
& \mu_0 = A_1B_0 - B_0A_0 + P_1Q_0 = 0,\\
& \mu_1 = A_0B_1 - B_1A_1 + P_0Q_1 = 0.
\end{aligned}
\end{equation*}

Let $G_{\mathbf k}:=GL(V_0)\times GL(V_1)$. We define the $\BZ_2$ chain-saw quiver variety by the GIT quotient
\[
    \mathcal{M}_{\mathbf k}^{\mathrm{cs}}
    :=
    \mu^{-1}(0)^{\mathrm{st}}/G_{\mathbf k},
\]
where the superscript \enquote{$\mathrm{st}$} indicates that we restrict to stable representations, meaning representations for which there is no proper pair of subspaces $S_\omega \subset V_\omega$ such that
\[
    A_\omega(S_\omega)\subset S_\omega,
    \qquad
    B_\omega(S_\omega)\subset S_{\omega+1},
    \qquad
    P_\omega(W_\omega)\subset S_\omega,
\]
for both $\omega=0,1$.

The torus acting on the chain-saw quiver variety is $T
    =
    \mathbb C^\times_{\varepsilon_1}
    \times
    \mathbb C^\times_{\hat{\varepsilon}_2}
    \times
    \mathbb C^\times_{a_0}
    \times
    \mathbb C^\times_{a_1}$. We write an element of $T$ as
\[
    t=(t_1,t_2,u_0,u_1),
    \qquad
    t_1=\re^{\varepsilon_1},\quad
    t_2=\re^{\hat{\varepsilon}_2},\quad
    u_\omega=\re^{a_{c^{-1}(\omega)}},
\]
where, for the $\mathrm{SU}(2)$ theory, we set $a_0=a$ and $a_1=-a$. Here $u_\omega$ acts on the one-dimensional framing space $W_\omega \cong \mathbb C$. The induced action on the chain-saw data is
\[
    A_\omega \mapsto t_1 A_\omega,
    \quad
    B_\omega \mapsto t_2 B_\omega, \quad  P_\omega \mapsto P_\omega u_\omega^{-1},
    \quad
    Q_\omega \mapsto t_1t_2\,u_{\omega+1}Q_\omega,
    \quad
    \omega\in \mathbb Z_2,
\]
where all indices are taken modulo $2$. This action preserves the moment map equations, since they transform as $\mu_\omega \longmapsto t_1t_2\,\mu_\omega$. Moreover, the $T$-action commutes with the action of $G_{\mathbf k}$, and hence descends to an action on $\mathcal M_{\mathbf k}^{\mathrm{cs}}$.

The $T$-fixed points $\left(\mathcal{M}_{\mathbf k}^{\mathrm{cs}}\right)^T$ are isolated and are classified by colored partitions. Since $\dim W_0=\dim W_1=1$, a fixed point is labeled by a pair of Young diagrams
\[
\boldsymbol{\lambda}
=
\bigl(\lambda^{(0)},\lambda^{(1)}\bigr).
\]
A box $(r,s)$ in $\lambda^{(\eta)}$, where $r$ and $s$ denote respectively its row and column indices, has $\mathbb Z_2$-weight $\eta+s-1\pmod 2$. Thus, odd columns have weight $\eta$, while even columns have weight $\eta+1\pmod 2$. The dimension vector $\mathbf k=(k_0,k_1)$ is recovered from the colored boxes by
\[
k_0
=
\sum_{\substack{s\geq1\\ s\ \mathrm{odd}}}
\lambda^{(0),t}_s
+
\sum_{\substack{s\geq1\\ s\ \mathrm{even}}}
\lambda^{(1),t}_s,
\qquad
k_1
=
\sum_{\substack{s\geq1\\ s\ \mathrm{even}}}
\lambda^{(0),t}_s
+
\sum_{\substack{s\geq1\\ s\ \mathrm{odd}}}
\lambda^{(1),t}_s.
\]
Thus
\[
    \left(\mathcal{M}_{\mathbf k}^{\mathrm{cs}}\right)^T
    =
    \left\{
    \bigl(\lambda^{(0)},\lambda^{(1)}\bigr)
    \ \middle|\
    |\boldsymbol\lambda|_0=k_0,\ 
    |\boldsymbol\lambda|_1=k_1
    \right\},
\]
where $|\boldsymbol\lambda|_\omega$ denotes the number of boxes of color $\omega$.

The equivariant Chern characters of the framing and tautological bundles, together with the characters of the flavor spaces, are then written as
\begin{subequations}
\begin{align}
&N_{\omega} = \re^{\,a_{c^{-1}(\omega)} - \omega\frac{\varepsilon_2}{2}},\\
&K_{\omega} = \sum_{\eta=0,1} \sum_{\substack{(r,s)\in\lambda^{(\eta)}\\
\eta+s-1\equiv\omega\!\!\!\!\pmod 2}}
\re^{a_{c^{-1}(\eta)}-\eta\frac{\varepsilon_2}{2}+\varepsilon_1(r-1)
+\frac{\varepsilon_2}{2}(s-1)},\\
& M_{0}^{(n_0)} = \sum_{\alpha=1}^{n_0}\re^{m_{2+\alpha}}, \qquad M_{1}^{(n_1)}= \sum_{\alpha=1}^{n_1}\re^{m_\alpha- \frac{\varepsilon_2}{2}},
\end{align}
\end{subequations}
where $\omega\in\mathbb{Z}_2$, and we denote by $N_f=(n_0,n_1)$ the distribution of fundamental hypermultiplets,\footnote{We are considering here a fixed choice of flavor coloring function $c_{\mathrm f}$ and splitting the masses accordingly.} $n_0,n_1\leq 2$.

In the preceding quiver description, $\hat{\varepsilon}_2$ denotes the equivariant weight of the covering coordinate $z_2$. In the characters above, we instead express the weights in terms of the rotation parameter $\varepsilon_2$ on the target space $\mathbb C^2$. Under the quotient map
\begin{equation}
    \mathbb C\times(\mathbb C/\mathbb Z_2)\longrightarrow\mathbb C^2,
    \qquad
    (z_1,z_2)\longmapsto
    (\tilde z_1=z_1,\tilde z_2=z_2^2),
\end{equation}
the target coordinate $\tilde z_2$ has equivariant weight $\varepsilon_2$, thus $\hat{\varepsilon}_2=\frac{\varepsilon_2}{2}$. This accounts for the replacement of the orbifold equivariant parameter by $\varepsilon_2/2$ in the characters above. Away from the ramification divisor $z_2=0$, a weight of $\BZ_2$-color $1$ can be converted into one of color $0$ by tensoring it with $z_2^{-1}$. In additive notation, this shifts its equivariant weight by
\begin{equation}
    w^{(1)}\longmapsto w^{(1)}-\frac{\varepsilon_2}{2}.
\end{equation}
This explains the color-dependent shifts $a_{c^{-1}(1)}\mapsto a_{c^{-1}(1)}-\varepsilon_2/2$ and $m_{\a}\mapsto m_{\a}-\varepsilon_2/2$ if $c_{\mathrm{f}} (\a) = 1$, while the color-$0$ weights remain unshifted. The trivialization by $z_2^{-1}$ is singular along $z_2=0$; consequently, smooth $\BZ_2$-equivariant fields on the covering space descend to fields with the prescribed surface singularity along $\widetilde z_2=0$.

The full equivariant character entering the localization formula is given by
\begin{equation}
\begin{aligned}
T[\bl]_{c}^{(n_0,n_1)}=&\sum_{\omega\in \mathbb{Z}_{2}}\bigg[N_{\omega}K_{\omega}^{*}+\re^{\varepsilon_1+\frac{\varepsilon_2}{2}}N_{\omega}^{*}K_{\omega-1}-(1-\re^{\varepsilon_1})K_{\omega}K_{\omega}^{*}\\
&+\re^{\frac{\varepsilon_2}{2}}(1-\re^{\varepsilon_1})K_{\omega}K_{\omega+1}^{*}-M_\omega^{(n_\omega)} K_\omega^*\bigg].
\end{aligned}
\end{equation}
The partition function of the orbifold theory is given by the generating function of the equivariant integrals of the Euler class of Dirac zero-mode bundle $\CalF$,
\begin{align}
\begin{split}
\Psi_{c,\,\text{inst}}^{(n_0,n_1)}\left(a,\boldsymbol{m},\varepsilon_1,\varepsilon_2;x,\mathfrak{q}\right)&= \sum_{k_0,k_1 =0} ^\infty \prod_{\o =0,1} \qe_\o ^{k_\o} \int_{\CalM_{\mathbf{k}} ^{\text{cs}}} e_{\mathsf{T}} (\mathcal{F}) \\
&=\sum_{\bl}\prod_{\omega=0,1}\mathfrak{q}^{k_{\omega}}_{\omega}\mathcal{E}\br{-T[\bl]_{c}^{(n_0,n_1)}}.
\end{split}
\end{align}
We denote the \textit{fractionalized} couplings by $(\mathfrak{q}_\omega)_{\o=0,1}$, which we choose to express in terms of the bulk gauge coupling $\qe$ and the monodromy defect parameter $x$ by
\begin{equation}
    \mathfrak{q}_0=\frac{\mathfrak{q}}{x}, \qquad \mathfrak{q}_1=x.
\end{equation}
Also, $k_{\omega}=\dim K_{\omega}$ and $\mathcal{E}$ denotes the operation that converts the equivariant Chern character into the equivariant Euler class,
\begin{equation}
    \mathcal{E}\left(\sum_{\alpha} \re^{w_{\alpha}} - \sum_{\beta} \re^{w_{\beta}} \right) = \frac{\prod_{\alpha} w_{\alpha}}{\prod_{\beta} w_{\beta}}.
\end{equation}
Finally, denoting by $\beta=0,1$ the two choices of coloring functions $c=\mathrm{id}$ and $c=(01)$, respectively, the full orbifold surface defect partition function takes the form
\begin{equation}
\label{eq: full orbifold}
\Psi_{\beta}^{(n_0,n_1)}\left(a,\boldsymbol{m},\varepsilon_1,\varepsilon_2;x,\mathfrak{q}\right)=\Psi_{\beta,\,\text{cl}}\,\Psi_{\beta,\,1\text{-loop}}^{(n_0,n_1)}\,\Psi_{\beta,\,\text{inst}}^{(n_0,n_1)},
\end{equation}
where the classical part of the partition function is
\begin{equation}
\Psi_{\beta,\,\text{cl}}\left(a,\varepsilon_1,\varepsilon_2;x,\mathfrak{q}\right) = \mathfrak{q}^{-\frac{a^2}{\varepsilon_1 \varepsilon_2}}x^{-\frac{a_{\beta}}{\varepsilon_1}},
\end{equation}
and the one-loop contribution is reported in Appendix \ref{app: special functions}. Analogous expressions for the bulk Nekrasov partition function $\mathcal{Z}^{(n_0,n_1)}$ can also be found in Appendix \ref{app: special functions}.

In the presence of a half-BPS surface defect, local observables in the defect and the bulk are subject to the chiral ring equation \cite{Gaiotto:2013sma}. In the presence of the $\O$-background, this chiral ring equation uplifts to a differential equation satisfied by the partition function \cite{Jeong:2018qpc,Nekrasov:2017rqy}. In particular, in the $N_f=(2,2)$ case, the orbifold partition functions \eqref{eq: full orbifold} can be shown to obey
\begin{equation}
\begin{aligned}
\label{eq: BPZ (2,2)}
&\left[\varepsilon_1^2\,
\partial_x^2 +\left(\frac{\varepsilon_1}{x}+\frac{m_1+m_2}{1-x}+\frac{\mathfrak{q}}{x}\frac{m_3+m_4}{ \mathfrak{q}-x}\right)\varepsilon_1\partial_x \right.\\
&\left.+\left(
\frac{m_1 m_2}{x(x-1)}
+\frac{\mathfrak{q}}{x}\frac{m_3 m_4}{x(\mathfrak{q}-x)}\right)+\frac{(\mathfrak{q}-1)\varepsilon_1\varepsilon_2}{x(x-\mathfrak{q})(x-1)} \mathfrak{q}\partial_\mathfrak{q}\right]\Psi_{\beta}^{(2,2)}=0
\end{aligned},\qquad \b=0,1.
\end{equation}
In the context of the AGT/BPS-CFT correspondence \cite{Alday:2009aq, Alday:2010vg, Nekrasov:2017gzb,Nekrasov:2021tik}, the regular orbifold surface-defect partition functions of the $\mathrm{SU}(2)$ $N_f = (2,2)$ theory are identified, up to elementary prefactors, with four-point conformal blocks of the affine Kac-Moody algebra $\widehat{\mathfrak{sl}}_2$. Accordingly, equation~\eqref{eq: BPZ (2,2)} can also be viewed as a gauge-transformed KZ equation satisfied by the affine conformal block. For $\widehat{\mathfrak{sl}}_2$, the KZ-BPZ correspondence expresses this four-point KZ conformal block as an integral transform of a five-point Virasoro conformal block on the sphere with one insertion of the level-two degenerate primary field $\phi_{2,1}(x)$, which solves the corresponding BPZ equation \cite{Ribault_2005,Frenkel:2015rda,Jeong:2024onv}. In this particular case, the reduced KZ equation and the BPZ equation admit the same form, with monodromy parameters shuffled by the integral transform.

In the Nekrasov--Shatashvili (NS) limit, defined by $\varepsilon_2\to 0$ and $\varepsilon_1=\hbar$, the bulk and defect partition functions behave as
\begin{subequations}
\label{eq: NS expansion}
\begin{align}
&\log\mathcal{Z}^{(n_0,n_1)}\simeq\frac{1}{\varepsilon_2}\mathcal{W}_0^{(n_0,n_1)}\left(a,\boldsymbol{m},\varepsilon_1;\mathfrak{q}\right)+\mathcal{W}_{1}^{(n_0,n_1)}\left(a,\boldsymbol{m},\varepsilon_1;\mathfrak{q}\right)+\mathcal{O}(\varepsilon_2),\\
&\log\Psi_{\beta}^{(n_0,n_1)}\simeq\frac{1}{\varepsilon_2}\mathcal{W}_0^{(n_0,n_1)}\left(a,\boldsymbol{m},\varepsilon_1;\mathfrak{q}\right)+\hat{\mathcal{W}}_{ \beta}^{(n_0,n_1)}\left(a,\boldsymbol{m},\varepsilon_1;x,\mathfrak{q}\right)+\mathcal{O}(\varepsilon_2),
\end{align}
\end{subequations}
where $\mathcal{W}_0^{(n_0,n_1)}\left(a,\boldsymbol{m},\varepsilon_1;\mathfrak{q}\right)$ coincides with the twisted superpotential of the effective two-dimensional $\mathcal{N}=(2,2)$ theory supported on the $z_2$-plane \cite{Nekrasov:2009rc}. Upon normalizing the defect partition function by the bulk one, the limit $\varepsilon_2 \to 0$ leads to a well-defined wavefunction,
\begin{align} \label{eq: normalized psi}
\begin{split}
\psi_\beta^{(n_0,n_1)}\left(a,\boldsymbol{m},\hbar;x,\mathfrak{q}\right)&:=\lim_{\varepsilon_2\to 0}\frac{\Psi_{\beta}^{(n_0,n_1)}\left(a,\boldsymbol{m},\hbar,\varepsilon_2;x,\mathfrak{q}\right)}{\mathcal{Z}^{(n_0,n_1)}\left(a,\boldsymbol{m},\hbar,\varepsilon_2;\mathfrak{q}\right)} \\&= \exp \left[ \hat{\mathcal{W}}_{ \beta}^{(n_0,n_1)}\left(a,\boldsymbol{m},\hbar;x,\mathfrak{q}\right) - \mathcal{W}_{1}^{(n_0,n_1)}\left(a,\boldsymbol{m},\hbar;\mathfrak{q}\right) \right].
\end{split}
\end{align}
In the $N_f=(2,2)$ case, this wavefunction satisfies the NS limit of \eqref{eq: BPZ (2,2)}, which is precisely the Heun equation,
\begin{equation}
\begin{aligned}
\label{eq: Heun (2,2)}
&\left[\hbar^2\,
\partial_x^2 +\left(\frac{\hbar}{x}+\frac{m_1+m_2}{1-x}+\frac{\mathfrak{q}}{x}\frac{m_3+m_4}{ (\mathfrak{q}-x)}\right)\hbar\partial_x \right.\\
&\left.+\left(
\frac{m_1 m_2}{x(x-1)}
+\frac{\mathfrak{q}}{x}\frac{m_3 m_4}{x(\mathfrak{q}-x)}\right)+\frac{(\mathfrak{q}-1)u^{(2,2)}}{x(x-\mathfrak{q})(x-1)} \right]\psi_{\beta}^{(2,2)}=0
\end{aligned},\qquad \b=0,1.
\end{equation}
By the monodromies of its solutions, the Heun operator \eqref{eq: Heun (2,2)} defines a parabolic $\mathrm{PGL}(2)$-local system on the four-punctured sphere $\BP^1\setminus \{0,\qe,1,\infty\}$. For generic values of the masses, the local monodromies around the four punctures are semisimple, with conjugacy classes fixed by the corresponding local exponents. For fixed $\mathfrak q$ and fixed local exponents, the space of such opers is an affine line, parametrized by the accessory parameter $u^{(2,2)}$. Through the relation
\begin{equation}
\label{eq: qmatone}
    u^{(2,2)}=\hbar\, \mathfrak{q}\partial_\mathfrak{q}\mathcal{W}_0^{(2,2)} = -a^2 + \hbar \, \mathfrak{q}\partial_\mathfrak{q}\mathcal{W}_{0,\, \text{inst}}^{(2,2)},
\end{equation}
the Coulomb parameter $a$ provides a local parametrization of this
oper line. Under the Riemann--Hilbert map, this oper locus forms a
holomorphic Lagrangian submanifold inside the moduli space of parabolic
$\mathrm{PGL}(2)$-local systems on the four-punctured sphere.

By \eqref{eq: Heun (2,2)}, the monodromy surface defect partition functions provide solutions to the Heun oper. Note that they are convergent in the domain $\vert \qe_0 \vert ,\vert \qe_1 \vert <1 $, which is equivalent to $ \vert  \qe\vert <\vert x \vert <1$. The two choices of the coloring function $\b\in \{0,1\}$ label the two independent solutions in this domain at generic values of the Coulomb parameter $a$, diagonal under the monodromy along the $A$-cycle of the four-punctured sphere $\BP^1 \setminus \{0,\qe,1,\infty\}$.

In this work, we depart from the generic values of the Coulomb parameter $a$ and study Heun solutions near the special loci $a \in \frac{\hbar}{2} \BZ$, where the naive series expansion \eqref{eq: normalized psi} of the wavefunction ceases to be well-defined. The method is the resummation of the defect partition functions near the special loci, which becomes available through the blow-up equations.

\subsection{Blow-up equations for surface defect partition functions}
It was physically argued and tested in \cite{Jeong:2020uxz} that surface defect partition functions satisfy blow-up equations, extending the corresponding results known for bulk partition functions \cite{Nakajima:2003pg}. See also \cite{Nekrasov:2020qcq}. In this work we will be interested in the blow-up equations obtained by considering the $\mathcal{N}=2$ gauge theory on $\hat{\mathbb{C}}^2$, the blow-up of $\mathbb{C}^2$ at the origin. The blow-up equations are labeled by the first Chern class $c_1 = k [E]$ of the torsion-free sheaf on $\hat{\mathbb{C}}^2$, with $k\in\{0,1\}$, where $E\hookrightarrow\hat{\mathbb{C}}^2$ denotes the exceptional divisor. Focusing once again on the $N_f=(2,2)$ orbifold partition functions, the blow-up equations corresponding to $k=0$ take the form
\begin{equation}
\label{eq: blowup 1}
\Psi_{\beta}^{(2,2)}(a,\boldsymbol{m},\varepsilon_1,\varepsilon_2;x,\mathfrak{q})=\sum_{n\in \mathbb{Z}}\Psi_\beta^{(2,2)}(a+n \varepsilon_1,\boldsymbol{m},\varepsilon_1,\varepsilon_2-\varepsilon_1;x,\mathfrak{q})\mathcal{Z}^{(2,2)}(a+n\varepsilon_2,\boldsymbol{m},\varepsilon_1-\varepsilon_2,\varepsilon_2;\mathfrak{q}),
\end{equation}
while the $k=1$ blow-up equations read
\begin{equation}
\label{eq: blowup 2 NF4}
\begin{aligned}
&\Psi_\beta^{(2,2)}(a,\boldsymbol{m},\varepsilon_1,\varepsilon_2;x,\mathfrak{q})=-\mathfrak{q}^{-\frac{1}{4}}x^{\frac{1}{2}}\frac{1-\mathfrak{q}}{1-x} \\
&\sum_{n\in \mathbb{Z}+\frac{1}{2}}\Psi_\beta^{(2,2)}\left(a+n \varepsilon_1,\boldsymbol{m}+\frac{\varepsilon_1}{2},\varepsilon_1,\varepsilon_2-\varepsilon_1;x,\mathfrak{q}\right)\mathcal{Z}^{(2,2)}\left(a+n\varepsilon_2,\boldsymbol{m}+\frac{\varepsilon_2}{2},\varepsilon_1-\varepsilon_2,\varepsilon_2;\mathfrak{q}\right).
\end{aligned}
\end{equation}
These equations have been proved using CFT methods and the AGT correspondence in \cite{Bershtein:2024kwe}. As discussed in \cite{Bonelli:2025bmt}, it is possible to take an appropriate limit of the blow-up equations that gives constraints on the NS asymptotic functions in equation \eqref{eq: NS expansion}. In particular, in the limit $\varepsilon_1=\varepsilon_2=\hbar$, the $k=0$ blow-up equations \eqref{eq: blowup 1} reduce to
\begin{equation}
\Psi_{\beta}^{(2,2)}(a,\boldsymbol{m},\hbar,\hbar;x,\mathfrak{q})=\sum_{n\in \mathbb{Z}}\exp\br{\hat{\mathcal{W}}_{\beta}^{(2,2)} + \mathcal{W}_{1}^{(2,2)} -\left(\partial_{\hbar}+ n \partial_{a}\right)\mathcal{W}_{0}^{(2,2)}},
\end{equation}
where $\hat{\mathcal{W}}_{\beta}^{(2,2)}=\hat{\mathcal{W}}^{(2,2)}_{\beta\,}\!\left(a+n\hbar,\boldsymbol{m},\hbar;x,\mathfrak{q}\right)$, and similarly for $\mathcal{W}_{0}^{(2,2)}$ and $\mathcal{W}_{1}^{(2,2)}$. Meanwhile, the $k=1$ blow-up equations \eqref{eq: blowup 2 NF4} reduce, in this limit, to
\begin{equation}
\begin{aligned}
\Psi_{\beta}^{(2,2)}(a,\boldsymbol{m},\hbar,\hbar;x,\mathfrak{q})&=-\mathfrak{q}^{-\frac{1}{4}}x^{\frac{1}{2}}\frac{1-\mathfrak{q}}{1-x}\\
&\times \sum_{n\in \mathbb{Z}+\frac{1}{2}}\exp\left( \hat{\mathcal{W}}_{\beta}^{(2,2)} + \mathcal{W}_{1}^{(2,2)} -\left(\frac{1}{2}\partial_{\boldsymbol{m}}+\partial_{\hbar}+ n \partial_{a}\right)\mathcal{W}_{0}^{(2,2)}\right),
\end{aligned}
\end{equation}
where $\partial_{\boldsymbol{m}}=\sum_i \partial_{m_i}$,
$\hat{\mathcal{W}}_{\beta}^{(2,2)}=\hat{\mathcal{W}}_{\beta\,}^{(2,2)}\!\left(a+n\hbar,\boldsymbol{m}+\hbar/2,\hbar;x,\mathfrak{q}\right)$, and similarly for $\mathcal{W}_{0}^{(2,2)}$ and $\mathcal{W}_{1}^{(2,2)}$. Equating the two right-hand sides then yields a blow-up equation involving only the first and second NS prepotentials:
\begin{equation}
\begin{aligned}
\label{eq: blowup combined (2,2)}
&\sum_{n\in \mathbb{Z}}\exp\br{\hat{\mathcal{W}}_{\beta}^{(2,2)} + \mathcal{W}_{1}^{(2,2)} -\left(\partial_{\hbar}+ n \partial_{a}\right)\mathcal{W}_{0}^{(2,2)}}\\
&+\mathfrak{q}^{-\frac{1}{4}}x^{\frac{1}{2}}\frac{1-\mathfrak{q}}{1-x}\sum_{n\in \mathbb{Z}+\frac{1}{2}}\exp\left( \hat{\mathcal{W}}_{\beta}^{(2,2)} + \mathcal{W}_{1}^{(2,2)} -\left(\frac{1}{2}\partial_{\boldsymbol{m}}+\partial_{\hbar}+ n \partial_{a}\right)\mathcal{W}_{0}^{(2,2)}\right)=0.
\end{aligned}
\end{equation}
This equation will be the main focus of the present work and will allow us to perform the resummation of the defect prepotentials $\hat{\mathcal{W}}_{\beta}^{(n_0,n_1)}$.

\subsection{Decoupling limits}
\label{sec: decoupling}
The differential and blow-up equations satisfied by the
$N_f=(n_0,n_1)$ partition functions $\Psi_\beta^{(n_0,n_1)}$ can be obtained by considering appropriate decoupling limits of the corresponding equations for the $N_f=(2,2)$ theory. Starting from equation \eqref{eq: BPZ (2,2)}, the decoupling of the masses $m_\alpha$, $\alpha\in\{1,2\}$, can be achieved by performing the substitutions $x\to x/m_\alpha$ and $\mathfrak{q}\to \mathfrak{q}/m_\alpha$, followed by the limit $m_\alpha\to\infty$. For instance, decoupling the mass $m_2$ leads to the following equation:
\begin{equation}
\left[\varepsilon_1^2\,
\partial_x^2 +\left(1+\frac{\mathfrak{q}}{x}\frac{m_3+m_4}{\mathfrak{q}-x}\right)\varepsilon_1\partial_x 
+\left(\frac{\mathfrak{q}}{x^2}\frac{m_3 m_4}{\mathfrak{q}-x}-
\frac{m_1}{x}\right)+\frac{\varepsilon_1\varepsilon_2}{x(x-\mathfrak{q})} \mathfrak{q}\partial_\mathfrak{q}\right]\Psi_{\beta}^{(2,1)}=0.
\end{equation}
Meanwhile, the decoupling of the masses $m_{2+\alpha}$, with
$\alpha\in\{1,2\}$, can be achieved by performing the substitution $\mathfrak{q}\to \mathfrak{q}/m_{2+\alpha}$, followed by the limit $m_{2+\alpha}\to\infty$. For instance, decoupling the mass $m_4$ leads to the equation
\begin{equation}
\left[\varepsilon_1^2\,
\partial_x^2 +\left(\frac{m_1+m_2}{1-x}-\frac{\mathfrak{q}}{x^2}\right)\varepsilon_1\partial_x 
+\left(
\frac{m_1 m_2}{x(x-1)}
-\frac{\mathfrak{q}}{x^3}m_3\right)+\frac{\varepsilon_1\varepsilon_2}{x^2(1-x)} \mathfrak{q}\partial_\mathfrak{q}\right]\Psi_{\beta}^{(1,2)}=0.
\end{equation}
In an analogous way, one can obtain all confluent degenerations of the Heun equation \eqref{eq: Heun (2,2)}. The same limits must also be implemented in the quantum Matone relation \eqref{eq: qmatone}, in order to recover the correct relation between the accessory parameter and the twisted effective superpotential of the theory obtained after hypermultiplet decoupling. 

Turning next to the blow-up equations, and performing the same limits on \eqref{eq: blowup 1} and \eqref{eq: blowup 2 NF4}, it is straightforward to see that the $k=0$ blow-up equations always take the form
\begin{equation}
\label{eq: blowup 1 (n0,n1)}
\begin{aligned}
&\Psi_{\beta}^{(n_0,n_1)}(a,\boldsymbol{m},\varepsilon_1,\varepsilon_2;x,\mathfrak{q})\\
&\quad =\sum_{n\in \mathbb{Z}}\Psi_\beta^{(n_0,n_1)}(a+n \varepsilon_1,\boldsymbol{m},\varepsilon_1,\varepsilon_2-\varepsilon_1;x,\mathfrak{q})\mathcal{Z}^{(n_0,n_1)}(a+n\varepsilon_2,\boldsymbol{m},\varepsilon_1-\varepsilon_2,\varepsilon_2;\mathfrak{q}),
\end{aligned}
\end{equation}
while in the $k=1$ case the blow-up equations read 
\begin{equation}
\begin{aligned}
&\Psi_\beta^{(n_0,n_1)}(a,\boldsymbol{m},\varepsilon_1,\varepsilon_2;x,\mathfrak{q})=c_{n_0,n_1}(x,\mathfrak{q})\times\\
&\sum_{n\in \mathbb{Z}+\frac{1}{2}}\Psi_\beta^{(n_0,n_1)}\left(a+n \varepsilon_1,\boldsymbol{m}+\frac{\varepsilon_1}{2},\varepsilon_1,\varepsilon_2-\varepsilon_1;x,\mathfrak{q}\right)\mathcal{Z}^{(n_0,n_1)}\left(a+n\varepsilon_2,\boldsymbol{m}+\frac{\varepsilon_2}{2},\varepsilon_1-\varepsilon_2,\varepsilon_2;\mathfrak{q}\right),
\end{aligned}
\end{equation}
where the prefactor $c_{n_0,n_1}$ is defined as
\begin{equation}
c_{n_0,n_1}(x,\mathfrak{q})=-\mathfrak{q}^{-\frac{1}{4}}x^{\frac{1}{2}}\cdot 
\begin{cases}
\frac{1-\mathfrak{q}}{1-x}, & \text{if } n_0=n_1 =2, \\[6pt]
\frac{1}{1-x}, & \text{if } n_0\leq 1 \text{ and } n_1 = 2, \\[6pt]
1, & \text{if } n_1 \leq 1 .
\end{cases}
\end{equation}
Finally, in the general $N_f=(n_0,n_1)$ case, the NS blow-up equations \eqref{eq: blowup combined (2,2)} take the form
\begin{equation}
\label{eq: blowup combined}
\begin{aligned}
&\sum_{n\in \mathbb{Z}}\exp\br{\hat{\mathcal{W}}_{\beta}^{(n_0,n_1)} + \mathcal{W}_{1}^{(n_0,n_1)} -\left(\partial_{\hbar}+ n \partial_{a}\right)\mathcal{W}_{0}^{(n_0,n_1)}}\\
&-c_{n_0,n_1}(x,\mathfrak{q})\sum_{n\in \mathbb{Z}+\frac{1}{2}}\exp\left( \hat{\mathcal{W}}_{\beta}^{(n_0,n_1)} + \mathcal{W}_{1}^{(n_0,n_1)} -\left(\frac{1}{2}\partial_{\boldsymbol{m}}+\partial_{\hbar}+ n \partial_{a}\right)\mathcal{W}_{0}^{(n_0,n_1)}\right)=0,
\end{aligned}
\end{equation}
where the dependence on the appropriately shifted arguments is always understood.

\section{Resummation of surface defect partition functions}
\label{sec:resummation}
In this section we begin by briefly reviewing the properties of the NS instanton functions $\mathcal{W}_{0,\,\mathrm{inst}}^{(n_0,n_1)}$, $\mathcal{W}_{1,\,\mathrm{inst}}^{(n_0,n_1)}$ and $\hat{\mathcal{W}}_{\beta,\,\mathrm{inst}}^{(n_0,n_1)}$, which motivate the resummation procedure and the form of the resummation ansatz that we will introduce \cite{Beccaria:2016wop, Gorsky:2017ndg, Bonelli:2025bmt}. 
To this end, we perform the convenient change of variables $x=z \mathfrak{t}$, $\mathfrak{q}=\mathfrak{t}^2$, in terms of which the Heun equation \eqref{eq: Heun (2,2)} takes the form
\begin{equation}
\label{eq: Heun (2,2) zt}
\begin{aligned}
&\left[\hbar^2\,
\partial_z^2 +\left(
\frac{\hbar}{z}
-\frac{\mathfrak{t}}{z}\frac{m_1+m_2}{\mathfrak{t}-z^{-1}}
+\frac{\mathfrak{t}}{z}\frac{m_3+m_4}{\mathfrak{t}-z}\right)\hbar\,\partial_z \right.\\
&\left.+\frac{\mathfrak{t}}{z^2}\left(
\frac{m_1 m_2}{\mathfrak{t}-z^{-1}}
+\frac{m_3 m_4}{\mathfrak{t}-z}\right)+\frac{(1-\mathfrak{t}^2)u^{(2,2)}}{z^2(\mathfrak{t}-z)(\mathfrak{t}-z^{-1})}\right]\psi_{\beta}^{(2,2)}=0,
\end{aligned}
\end{equation}
and where the accessory parameter is given by
\begin{equation}
\label{eq:acc_param_zt}
u^{(2,2)}= 
\frac{\hbar}{2}\mathfrak{t}\,\partial_\mathfrak{t} \mathcal{W}_{0}^{(2,2)}.
\end{equation}
With a slight abuse of notation, we will denote
\begin{equation}
\mathcal{W}_{i,\,\text{inst}}^{(n_0,n_1)}
\br{a,\boldsymbol{m},\hbar;\mathfrak{t}^2}
\equiv
\mathcal{W}_{i,\,\text{inst}}^{(n_0,n_1)}
\br{a,\boldsymbol{m},\hbar;\mathfrak{t}}, \qquad i = 0,1,
\end{equation}
and, similarly,
\begin{equation}
\hat{\mathcal{W}}_{\beta,\,\text{inst}}^{(n_0,n_1)}
\br{a,\boldsymbol{m},\hbar;z\mathfrak{t},\mathfrak{t}^2}
\equiv
\hat{\mathcal{W}}_{\beta,\,\text{inst}}^{(n_0,n_1)}
\br{a,\boldsymbol{m},\hbar;z,\mathfrak{t}}, \qquad \beta = 0,1.
\end{equation}
The intended meaning will always be clear from the variables appearing in the arguments. We then proceed to show how the blow-up equations \eqref{eq: blowup combined} can be used to determine exactly the resummation functions parametrizing the NS instanton function $\hat{\mathcal{W}}_{\beta,\,\mathrm{inst}}^{(n_0,n_1)}$, using as input the resummation functions parametrizing $\mathcal{W}_{0,\,\mathrm{inst}}^{(n_0,n_1)}$ and $\mathcal{W}_{1,\,\mathrm{inst}}^{(n_0,n_1)}$. The latter can in turn be determined by following the similar procedure described in \cite{Bonelli:2025bmt}. We then discuss the resulting pattern that emerges for the resummation functions, their analytic structure, and how these properties allow one to safely take the limit to half-integer values of $a/\hbar$, correctly reproducing the expected wavefunctions. Finally, we explain how the logarithmic companions of the resonant solutions can be obtained from the resummed wavefunctions, discuss the semisimple locus of the resonant $A$-cycle monodromy, and conclude with the explicit example of the $N_f=(1,1)$ theory.

\subsection{Resummation ansatz}
In the following discussion we suppress the numbers of hypermultiplets $n_0$ and $n_1$, and set $\hbar=1$. The dependence on $\hbar$ can be reinstated by rescaling the remaining variables and the NS functions according to
\begin{subequations}
\label{eq:hbar_rescalings}
\begin{align}
&\mathcal{W}_{0,\,\text{inst}}^{(n_0,n_1)}(a,\boldsymbol{m},\hbar;\mathfrak{t})=\hbar\,\mathcal{W}_{0,\,\text{inst}}^{(n_0,n_1)}\br{a\hbar^{-1},\boldsymbol{m}\hbar^{-1},1;\mathfrak{t}\,\hbar^{\frac{n_0+n_1}{2}-2}},\\
&\mathcal{W}_{1,\,\text{inst}}^{(n_0,n_1)}(a,\boldsymbol{m},\hbar;\mathfrak{t})=\mathcal{W}_{1,\,\text{inst}}^{(n_0,n_1)}\br{a\hbar^{-1},\boldsymbol{m}\hbar^{-1},1;\mathfrak{t}\,\hbar^{\frac{n_0+n_1}{2}-2}},\\
&\hat{\mathcal{W}}_{\beta,\,\text{inst}}^{(n_0,n_1)}(a,\boldsymbol{m},\hbar;z,\mathfrak{t})=\hat{\mathcal{W}}_{\beta,\,\text{inst}}^{(n_0,n_1)}\br{a\hbar^{-1},\boldsymbol{m}\hbar^{-1},1;z\, \hbar^{\frac{n_1-n_0}{2}},\mathfrak{t}\,\hbar^{\frac{n_0+n_1}{2}-2}}.
\end{align}
\end{subequations}

\paragraph{$\boldsymbol{\mathcal{W}_{0,\,\text{inst}}}$.} The instanton prepotential $\mathcal{W}_{0,\,\text{inst}}(a,\boldsymbol{m},\hbar;\mathfrak{t})$
is a symmetric function of the masses $\boldsymbol{m}$ and an even function of both $a$ and $\mathfrak{t}$. When expanded around $\mathfrak{t}=0$, it admits the power-series representation
\begin{equation}
\mathcal{W}_{0,\,\text{inst}}(a,\boldsymbol{m},1;\mathfrak{t})=R_0(a,\boldsymbol{m},1;\mathfrak{t})+\sum_{k=1}^{\infty}f_{0,k}(a,\boldsymbol{m})\mathfrak{t}^{2k},
\end{equation}
where the coefficients $f_{0,k}$ are rational functions of the remaining variables, with a predictable pole structure, namely with fixed locations and degrees of the poles at each order in the expansion, while $R_0$ denotes the part of the prepotential regular in $a$, see equation \eqref{eq: regular parts}. In particular, upon decomposition into simple fractions, each function $f_{0,k}$ takes the form
\begin{equation}
f_{0,k}(a,\boldsymbol{m})=
\sum_{j=1}^{k}\ \sum_{p=1}^{\left\lfloor k/j \right\rfloor}
\sum_{\pm}\!
\frac{d^{(k)}_{j,p}(\boldsymbol{m})}{(j\pm 2a)^{\,2\left\lfloor k/j \right\rfloor-2p+1}},
\end{equation}
where the coefficients $d^{(k)}_{j,p}(\boldsymbol{m})$ are symmetric polynomials in the masses $\boldsymbol{m}$ with rational coefficients. It is also worth noting that $a=0$ is a regular point of the instanton prepotential. This structure naturally leads to the resummation ansatz
\begin{equation}
\label{eq: W0 resum}
\mathcal{W}_{0,\,\text{inst}}(a,\boldsymbol{m},1,\mathfrak{t})=R_0(a,\boldsymbol{m},1;\mathfrak{t})+\sum_{k,j=1}^{\infty}\sum_{\pm}\mathfrak{t}^{2k+j-2}g_{k,j}\br{\boldsymbol{m},\frac{\mathfrak{t}^{j}}{j\pm2a}},
\end{equation}
where the functions\footnote{In this section, the variable $x$ denotes the last argument of the resummation functions under consideration and should not be confused with the variable appearing in the Heun equation \eqref{eq: Heun (2,2)}, which is replaced throughout by $z\mathfrak{t}$.} $g_{k,j}(\boldsymbol{m},x)$ are odd and analytic in a neighborhood of $x=0$, so as to reproduce the correct pole structure of the standard instanton expansion. The general structure that seems to emerge for the resummation functions is the following \cite{Bonelli:2025bmt}:
\begin{subequations}
\label{eq: W0 res fun}
\begin{align}
& g_{1,j}(\boldsymbol{m},x)=-\frac{1}{x}\left(\log\!\left[\frac{1}{2}+\frac{1}{2}\sqrt{1+4 w_j^{(n_0,n_1)}x^2/\zeta_j^2} \right]+1-\sqrt{1+4 w_j^{(n_0,n_1)}x^2/\zeta_j^2}\right),\\
& g_{k\geq2,j}(\boldsymbol{m},x) =\frac{1}{x^{2k-1}}\left[\br{1+4 w_j^{(n_0,n_1)}x^2/\zeta_j^2}^{\frac{5}{2}-k}Q_{k,j}(\boldsymbol{m},x^2)+P_{k,j}(\boldsymbol{m},x^2) \right],
\end{align}
\end{subequations}
where $Q_{k,j}$ and $P_{k,j}$ are polynomials in $x^2$ of degrees $2k-3$ and $k-1$, respectively,
\begin{equation}
    Q_{k,j}(\boldsymbol{m},x^2)=\sum_{\alpha=0}^{2k-3}q_{k,j,\alpha}(\boldsymbol{m})x^{2\alpha}, \qquad P_{k,j}(\boldsymbol{m},x^2)=\sum_{\alpha=0}^{k-1}p_{k,j,\alpha}(\boldsymbol{m})x^{2\alpha},
\end{equation}
while 
\begin{subequations}
\begin{align}
&w_j^{(n_0,n_1)}=\prod_{k=1-\frac{j}{2}}^{\frac{j}{2}}\prod_{i=1}^{n_0}(m_{i+2}+k)\prod_{i=1}^{n_1}(m_{i}+k),\\
&\zeta_j=(-1)^{j-1}j!(j-1)! .
\end{align}
\end{subequations}
The polynomials $P_{k,j}$ are not independent of the polynomials $Q_{k,j}$, since the coefficients $p_{k,j,\alpha}$ are fixed in terms of the $q_{k,j,\alpha}$ by imposing the regularity condition
\begin{equation}
g_{k,j}(\boldsymbol{m},x)=\mathcal{O}(x), \qquad x \to 0.
\end{equation}
The logarithmic structure appearing in the leading resummation functions $g_{1,j}$ is crucial for correcting the perturbative one-loop contribution to the twisted effective superpotential, as shown in \cite{Gorsky:2017ndg}.

Finally, although $x=\infty$ is a branch point of the functions $g_{k,j}$, these functions remain finite as $x\to\pm\infty$. This makes it possible to define a resummed accessory parameter $u_{\mathrm{res}}$ through the relation \eqref{eq:acc_param_zt}, using the resummed expression \eqref{eq: W0 resum} for the instanton contribution. Then, the resulting $u_{\mathrm{res}}$ remains well defined in the resonant limits $2a\to\pm n$, with $n\in\mathbb{N}$. Because of the branch structure introduced by the resummation functions, the value of the accessory parameter at resonance depends on the direction from which the limit is approached. More precisely, one obtains the two branches,
\begin{equation}
    u^{(n)}_{\pm}=\lim_{2a\to n^{\pm}} u_{\mathrm{res}} = \lim_{2a\to (-n)^{\mp}} u_{\mathrm{res}}, \qquad n \in \mathbb{N}.
\end{equation}
The resummed accessory parameter can then be used to construct an infinite family of functions of a quasi-momentum $\kappa$:
\begin{equation}
\label{eq:dispersions_bands}
    E^{(n)}\br{\kappa,\boldsymbol{m},\hbar;\mathfrak{t}} =
    \begin{cases}
        -u_{\mathrm{res}}\br{\frac{n}{2}+\kappa,\boldsymbol{m},\hbar;\mathfrak{t}}, & n\geq0 \text{ even},\\
        -u_{\mathrm{res}}\br{\frac{n+1}{2}-\kappa,\boldsymbol{m},\hbar;\mathfrak{t}}, & n>0 \text{ odd},
    \end{cases}
\end{equation}
initially defined for $\kappa\in\left[0,1/2\right]$. Each function $E^{(n)}$ is first extended to the interval $\kappa\in\left[-1/2,1/2\right]$ by imposing the reflection symmetry
\begin{equation}
E^{(n)}\br{-\kappa,\boldsymbol{m},\hbar;\mathfrak{t}}
=
E^{(n)}\br{\kappa,\boldsymbol{m},\hbar;\mathfrak{t}}.
\end{equation}
The branches reached at the endpoints $\kappa=-\frac{1}{2}$ and $\kappa=\frac{1}{2}$ are then identified, and this gluing is repeated along the real $\kappa$-axis. In this way, each $E^{(n)}$ is extended to an even, $1$-periodic function of the quasi-momentum $\kappa$. The resulting $E^{(n)}$ can then be interpreted as the dispersion relations describing the spectral bands of the periodic spectral problems\footnote{Under the change of variables $z=\re^{2\pi\ri y}$, the Heun equation \eqref{eq: Heun (2,2) zt} becomes a differential equation with coefficients that are periodic in $y$ with period one.} associated with the Heun equation \eqref{eq: Heun (2,2) zt} and its confluent limits.

\paragraph{$\boldsymbol{\mathcal{W}_{1,\,\text{inst}}}$.} The prepotential $\mathcal{W}_{1,\,\text{inst}}(a,\boldsymbol{m},\hbar;\mathfrak{t})$ is also a symmetric function of the masses $\boldsymbol{m}$ and an even function of $a$ and $\mathfrak{t}$, admitting the power-series representation
\begin{equation}
\mathcal{W}_{1,\,\text{inst}}(a,\boldsymbol{m},1;\mathfrak{t})=R_1(a,\boldsymbol{m},1;\mathfrak{t})+\sum_{k=1}^{\infty}f_{1,k}(a,\boldsymbol{m})\mathfrak{t}^{2k},
\end{equation}
where the coefficients $f_{1,k}$ are again rational functions with a predictable pole structure, and $R_1$ denotes the part of $\mathcal{W}_{1,\,\text{inst}}$ that is regular in $a$, see equation \eqref{eq: regular parts}. Upon decomposing the functions $f_{1,k}$ into simple fractions, we find
\begin{equation}
f_{1,k}(a,\boldsymbol{m}) =
\sum_{j=1}^{k}\ \sum_{p=1}^{\left\lfloor k/j \right\rfloor}\sum_{\pm}
\!\left(
\frac{\alpha^{(k)}_{j,p}(\boldsymbol{m})}{(j\pm2a)^{\,2\left\lfloor k/j \right\rfloor-2p+2}} + \frac{\beta^{(k)}_{j,p}(\boldsymbol{m})}{(j\pm2a)^{\,2\left\lfloor k/j \right\rfloor-2p+1}}
\right),
\end{equation}
where $\alpha^{(k)}_{j,p}(\boldsymbol{m})$ and $\beta^{(k)}_{j,p}(\boldsymbol{m})$ are symmetric polynomials in the masses $\boldsymbol{m}$ with rational coefficients. Again, it is worth noting that $a=0$ is a regular point of $\mathcal{W}_{1,\,\text{inst}}$. In contrast to $\mathcal{W}_{0,\,\text{inst}}$, the pole structure of each coefficient now involves both even and odd powers. This motivates the introduction of two distinct sets of resummation functions in the ansatz
\begin{equation}
\label{eq: W1 resum}
\mathcal{W}_{1,\,\text{inst}}(a,\boldsymbol{m},1,\mathfrak{t})=R_1(a,\boldsymbol{m},1;\mathfrak{t})+\sum_{k,j=1}^{\infty}\sum_{\pm}\mathfrak{t}^{2k-2}\left[\tilde{f}_{k,j}\br{\boldsymbol{m},\frac{\mathfrak{t}^j}{j\pm 2a}}+\mathfrak{t}^{j}\tilde{g}_{k,j}\br{\boldsymbol{m},\frac{\mathfrak{t}^j}{j\pm 2a}} \right]
\end{equation}
where the functions $\tilde{f}_{k,j}(\boldsymbol{m},x)$ and
$\tilde{g}_{k,j}(\boldsymbol{m},x)$ are analytic and vanishing in a neighborhood of $x=0$, and are even and odd in $x$, respectively. Their explicit form appears to follow the pattern
\begin{equation}
\label{eq: g-tilde functions}
\tilde{g}_{k,j}(\boldsymbol{m},x)
=
\frac{1}{x^{2k-1}}
\left[
\left(1+\frac{4 w_j^{(n_0,n_1)} x^2}{\zeta_j^2}\right)^{\frac{3}{2}-k}
U_{k,j}(\boldsymbol{m},x^2)
+
V_{k,j}(\boldsymbol{m},x^2)
\right],
\end{equation}
for the $\tilde g$-functions, while for the $\tilde f$-functions we find
\begin{equation}
\label{eq: f-tilde functions}
\tilde{f}_{1,j}(\boldsymbol{m},x)
=
-\frac{1}{4}\log\!\left(1+\frac{4 w_j^{(n_0,n_1)} x^2}{\zeta_j^2}\right),
\qquad
\tilde{f}_{k\geq 2,j}(\boldsymbol{m},x)
=
\frac{x^2\, Z_{k,j}(\boldsymbol{m},x^2)}
{\left(1+4 w_j^{(n_0,n_1)} x^2/\zeta_j^2\right)^{k-1}}.
\end{equation}
The functions $U_{k,j}$, $V_{k,j}$, and $Z_{k,j}$ are polynomials of degrees $2k-2$, $k-1$, and $k-2$, respectively\footnote{The coefficients of the polynomials $U_{k,j}$, $V_{k,j}$, and $Z_{k,j}$ differ from those in \cite{Bonelli:2025bmt} because we use a different convention for the fundamental mass parameters.}. Moreover, as in the case of the NS function $\mathcal{W}_{0,\,\mathrm{inst}}$, the polynomials $V_{k,j}$ are not independent of the polynomials $U_{k,j}$, but are fixed by imposing
\begin{equation}
\tilde{g}_{k,j}(\boldsymbol{m},x)=\mathcal{O}(x), \qquad x \to 0.
\end{equation}
As one approaches a pole in $a$, the resummation functions $\tilde{g}_{k,j}$ above develop a branch-point non-analyticity while remaining finite. Similarly, the functions $\tilde f_{k,j}$ remain finite with the exception of $\tilde{f}_{1,j}$, whose singular part behaves as
\begin{equation}
\tilde{f}_{1,j}\!\left(\boldsymbol{m},\frac{\mathfrak{t}^j}{j\pm 2a}\right)
=
-\frac{1}{4}\log \frac{\mathfrak{t}^{2j}}{(j\pm 2a)^2} + \mathcal{O}(1),
\qquad 2a\to \mp j.
\end{equation}
This singular behavior can be canceled by adding to the NS prepotential the combination
\begin{equation}
\frac{1}{2}\log\!\bigl(\Gamma(1+2a)\Gamma(1-2a)\bigr)
=
-\frac{1}{2}\log(2a\pm j) + \mathcal{O}(1),
\qquad 2a\to \mp j,
\end{equation}
which precisely compensates the singular contribution coming from $\tilde{f}_{1,j}$.

\paragraph{ $\boldsymbol{\hat{\mathcal{W}}_{\beta,\,\text{inst}}}$.} The defect instanton prepotentials $\hat{\mathcal{W}}_{\beta,\,\text{inst}}(a,\boldsymbol{m},\hbar;z,\mathfrak{t})$ are symmetric functions of the two sets of masses $\{m_\alpha\}$ and $\{m_{2+\alpha}\}$, with $\alpha\in\{1,2\}$, independently, namely they are invariant under permutations within each set, but not under permutations mixing the two sets. Since the two defect instanton functions are related by
\begin{equation}
    \hat{\mathcal{W}}_{1,\,\text{inst}}(a,\boldsymbol{m},\hbar,z,\mathfrak{t})=\hat{\mathcal{W}}_{0,\,\text{inst}}(-a,\boldsymbol{m},\hbar,z,\mathfrak{t}),
\end{equation}
we can focus on the $\beta=0$ case. The NS function admits the power-series representation
\begin{equation}
\hat{\mathcal{W}}_{0,\,\text{inst}}(a,\boldsymbol{m},1;z,\mathfrak{t})=\hat{R}_0(a,\boldsymbol{m},1;z,\mathfrak{t})+\sum_{k=1}^{\infty}\mathfrak{f}_{0,k}(a,\boldsymbol{m};z)\mathfrak{t}^{k},
\end{equation}
where the coefficients $\mathfrak{f}_{0,k}$ are once again rational functions, while $\hat{R}_0$ denotes the part of the function that is regular in $a$, see equation \eqref{eq: regular parts}. The functions $\mathfrak{f}_{0,k}$ admit the simple-fraction decomposition
\begin{equation}
\mathfrak{f}_{0,k}(a,\boldsymbol{m};z) =
\sum_{j=1}^{k}\ \sum_{p=1}^{\left\lfloor k/j \right\rfloor}
\!\left(
\frac{\hat{\alpha}^{(k)}_{j,p}(\boldsymbol{m},z)}{(j+2a)^{\,\left\lfloor k/j \right\rfloor-p+1}} + \frac{\hat{\beta}^{(k)}_{j,p}(\boldsymbol{m},z)}{(j-2a)^{\,\left\lfloor k/j \right\rfloor-p+1}}
\right),
\end{equation}
where the coefficients $\hat{\alpha}^{(k)}_{j,p}(\boldsymbol{m},z)$ and $\hat{\beta}^{(k)}_{j,p}(\boldsymbol{m},z)$ are symmetric functions of the two sets of masses $\{m_\alpha\}$ and $\{m_{2+\alpha}\}$, with $\alpha\in\{1,2\}$, independently. As in the case of the NS functions discussed above, the defect NS prepotential $\hat{\mathcal{W}}_{0,\,\text{inst}}$ is also regular at $a=0$. Given the pole structure discussed above, we can consider a resummation ansatz involving two families of functions:
\begin{equation}
\label{eq: What resum}
\hat{\mathcal{W}}_{0,\,\text{inst}}\left(a,\boldsymbol{m},1;z,\mathfrak{t}\right)=\hat{R}_0(a,\boldsymbol{m},1;z,\mathfrak{t})+\sum_{k,j=1}^{\infty}\sum_{\pm}\mathfrak{t}^{k-1}h^{\pm}_{k,j}\br{\boldsymbol{m},z,\frac{\mathfrak{t}^j}{j\pm 2a}},
\end{equation}
such that the functions $h^{\pm}_{k,j}(\boldsymbol{m},z,x)$ are analytic and vanish in a neighborhood of $x=0$. Alternatively, the form of the blow-up equation \eqref{eq: blowup combined} suggests incorporating $\mathcal{W}_{1,\,\text{inst}}$ into $\hat{\mathcal{W}}_{\beta,\,\text{inst}}$, which, from the perspective of the Heun equation \eqref{eq: Heun (2,2) zt}, amounts to a change of wavefunction normalization and leads to the following resummation ansatz:
\begin{equation}
\label{eq: What + W1 resum}
\begin{aligned}
&\hat{\mathcal{W}}_{0,\,\text{inst}}\left(a,\boldsymbol{m},1;z,\mathfrak{t}\right)+\mathcal{W}_{1,\,\text{inst}}\left(a,\boldsymbol{m},1;\mathfrak{t}\right)\\
&=\hat{R}_0(a,\boldsymbol{m},1;z,\mathfrak{t})+R_1(a,\boldsymbol{m},1;\mathfrak{t})+\sum_{k,j=1}^{\infty}\sum_{\pm}\mathfrak{t}^{k-1}\tilde{h}^{\pm}_{k,j}\br{\boldsymbol{m},z,\frac{\mathfrak{t}^j}{j\pm 2a}},
\end{aligned}
\end{equation}
where the functions $\tilde{h}^{\pm}_{k,j}(\boldsymbol{m},z,x)$ are also analytic and vanish in a neighborhood of $x=0$. The two families of resummation functions are then related by
\begin{equation}
\label{eq: relation h-funs}
   \tilde{h}_{k,j}^{\pm}(\boldsymbol{m},z,x)=h_{k,j}^{\pm}(\boldsymbol{m},z,x)+\sum_{r\geq 1}\left( \delta_{k,2r-1}\tilde{f}_{r,j}(\boldsymbol{m},x)+\delta_{k,2r+j-1}\tilde{g}_{r,j}(\boldsymbol{m},x)\right).
\end{equation}
\subsection{Resummation procedure}
The blow-up equations \eqref{eq: blowup combined} allow one to determine the resummation functions $h^{\pm}_{k,j}$ in equation \eqref{eq: What resum}, provided that the resummation functions $g_{k,j}$, $\tilde{g}_{k,j}$ and $\tilde{f}_{k,j}$ for the NS prepotentials $\mathcal{W}_0$ and $\mathcal{W}_1$ are known. The latter can be derived from the standard $\hat{\mathbb{C}}^2$ blow-up equations satisfied by the bulk partition function $\mathcal{Z}$ \cite{Bonelli:2025bmt}. Meanwhile, to determine the resummation functions $\tilde{h}^{\pm}_{k,j}$ in equation \eqref{eq: What + W1 resum}, only the functions $g_{k,j}$ are necessary. Since the equations \eqref{eq: blowup combined} for $\beta=0$ and $\beta=1$ are related by the transformation $a\to -a$, together with $n\to -n$, we restrict our analysis to the $\beta=0$ case. By suppressing the \enquote{instanton} label, setting $\hbar=1$, and applying the substitutions $\mathfrak{q}=\mathfrak{t}^2$ and $x = z\,\mathfrak{t}$, the blow-up equation \eqref{eq: blowup combined} can be rewritten as
\begin{equation}
\begin{aligned}
\label{eq: blowup res}
&0=\sum_{n \in \mathbb{Z}}\mathfrak{t}^{n (2 n-1)} z^{-n}\ell_{n}^{(n_0,n_1)}(a,\boldsymbol{m})\,\re^{\hat{\mathcal{W}}_0+\mathcal{W}_1-\left(\partial_\hbar+n\partial_a\right)\mathcal{W}_0}\\
&-c_{n_0,n_1}(z \mathfrak{t},\mathfrak{t}^2)\sum_{n \in \mathbb{Z}+\frac{1}{2}}\mathfrak{t}^{n (2 n-1)} z^{\frac{1}{2}-n}\ell_{n}^{(n_0,n_1)}(a,\boldsymbol{m})\,\re^{\hat{\mathcal{W}}_0+\mathcal{W}_1-\left(\frac{1}{2}\sum_{i}\partial_{m_i}+\partial_\hbar+n\partial_a\right)\mathcal{W}_0},
\end{aligned}
\end{equation}
where the notation $\hat{\mathcal{W}}_{0}=\hat{\mathcal{W}}_{0}(a+n, \boldsymbol{m},1;z,\mathfrak{t})$ is understood, and similarly for $\mathcal{W}_{0}$ and $\mathcal{W}_{1}$. The $\ell$-factors can be written as
\begin{equation}
\label{eq: ell factors}
    \ell_n^{(n_0,n_1)}(a,\boldsymbol{m})=g_{n}^{v}(a)\hat{g}_{n}^{v}(a)\prod_{i_0=1}^{n_0}g_n^f(a,m_{i_0+2})\prod_{i_1=1}^{n_1}g_n^f(a,m_{i_1})\hat{g}_n^{f}(a,m_{i_1}), \quad n \in \frac{1}{2}\mathbb{Z},
\end{equation}
where the various factors appearing in the product are given by
\begin{subequations}
\label{eq: ell factors explicit}
\begin{align}
& g_n^{ v}(a)=
\begin{cases}
\prod_{j=1}^{2n}\,(-2a-j+1)^{-j}\,(2a+j)^{\,1-j},
& n\ge 0,\\[0.6em]
g_{-n}^{ v}(-a),
& n<0~,
\end{cases}\\[0.6em]
&g_n^{ f}(a,m)=
\begin{cases}
\prod_{j_0=0}^{\lfloor n\rfloor}(-a-j_0+m+1)^{j_0}
\prod_{j_1=1}^{\left\lfloor n+\frac12\right\rfloor}(a+j_1+m)^{j_1-1},
& n\ge 0,\\[0.6em]
g_{-n}^{ f}(-a,m),
& n<0~,
\end{cases}\\[0.6em]
&\hat{g}_n^{v}(a)=\frac{\Gamma(2a+2n)}{\Gamma(2a)}, \qquad \qquad \hat{g}_n^{f}(a,m)=\frac{\Gamma(a-m)}{\Gamma(a-m+n-\xi_n)}.
\end{align}
\end{subequations}
The constant $\xi_n$ appearing in the last equation vanishes for $n \in \mathbb{Z}$, while $\xi_n = 1/2$ for $n \in \mathbb{Z}+1/2$. To derive from \eqref{eq: blowup res} the equations determining the resummation functions $h^{\pm}_{k,j}$, we set
\begin{equation}
    a = \frac{1}{2}\left(\sigma_1+\frac{\mathfrak{t}^{\sigma_2}}{x} \right),\quad \sigma_i\in \mathbb{Z}, \quad \sigma_2>0,
\end{equation}
where $x$ is a complex variable, and expand \eqref{eq: blowup res} around $\mathfrak{t}=0$. This yields a series expansion in $\mathfrak{t}$, which may start at a positive power. By setting the coefficients of this expansion to zero, starting from the first non-vanishing term, \eqref{eq: blowup res} provides algebraic equations, linear once we set $h^{\pm}_{1,\,j}=\log H^{\pm}_{1,\,j}$,  constraining
$h^{+}_{k,\sigma_2}(\boldsymbol{m},z,x)$ and $h^{-}_{k,\sigma_2}(\boldsymbol{m},z,x)$. Since the equations obtained for different values of $\sigma_1$ are independent, they can be straightforwardly solved for $h^{\pm}_{k,\sigma_2}$. However, determining the functions at level $j=\sigma_2+1$ requires knowledge of both the functions at level $j=\sigma_2$ and the functions $g_{k,j}$, $\tilde{g}_{k,j}$ and $\tilde{f}_{k,j}$ at level $j=\sigma_2+1$. Consequently, we proceed iteratively: we first compute the resummation functions for $\sigma_2=1$ up to a fixed value of $k$, then we proceed to $\sigma_2=2$, and so forth. A concrete example of this procedure is presented in Appendix~\ref{app: example}, where we write down and solve the first few equations arising from the expansion of \eqref{eq: blowup res} for the $N_f=(1,1)$ case.

\subsubsection{Leading order resummation functions \& split eigenfunctions}
\label{sec: leading order}

The previous resummation procedure is computationally demanding in the most general $N_f=(2,2)$ case, especially when the mass parameters are kept generic, as we assume throughout this subsection. In this case, we obtained only the first few $j=1$ and $j=2$ resummation functions. In theories with fewer hypermultiplets, it is easier to reach higher orders, particularly in the $N_f=(0,0)$ case. The $N_f=(1,1)$ theory provides a useful intermediate case for understanding the structure of the general resummation functions at higher orders, since it retains much of the structure of the $N_f=(2,2)$ case while being computationally less demanding. Guided by the explicit form of the first few leading resummation functions in these examples, we conjecture the following general structure for the leading resummation functions of the first family appearing in \eqref{eq: What resum}, for arbitrary $j$:
\begin{subequations}
\label{eq: resumm leading h}
\begin{align}
&h_{1,j}^{+}(\boldsymbol{m},z,x)=\log \left(\frac{\sqrt{1+4 w_j^{(n_0,n_1)}x^2/\zeta_j^2}+2 z^j v_{1,\,j}^{(n_1)} x/\zeta_j-1}{2z^j v_{1,\,j}^{(n_1)} x/\zeta_j\br{1+4 w_j^{(n_0,n_1)}x^2/\zeta_j^2}^{\frac{1}{4}}}\right),\\
&h_{1,j}^{-}(\boldsymbol{m},z,x)=\log \left(\frac{\sqrt{1+4 w_j^{(n_0,n_1)}x^2/\zeta_j^2}+2 z^j v_{1,\,j}^{(n_1)} x/\zeta_j+1}{2 \br{1+4 w_j^{(n_0,n_1)}x^2/\zeta_j^2}^{\frac{1}{4}}}\right).
\end{align}
\end{subequations}
Analogously, for the second family in \eqref{eq: What + W1 resum}, we find
\begin{subequations}
\label{eq: resumm leading ht}
\begin{align}
&\tilde{h}_{1,j}^{+}(\boldsymbol{m},z,x)=\log \left(\frac{\sqrt{1+4 w_j^{(n_0,n_1)}x^2/\zeta_j^2}+2 z^j v_{1,\,j}^{(n_1)} x/\zeta_j-1}{2z^j v_{1,\,j}^{(n_1)} x/\zeta_j\sqrt{1+4 w_j^{(n_0,n_1)}x^2/\zeta_j^2}}\right),\\
&\tilde{h}_{1,j}^{-}(\boldsymbol{m},z,x)=\log \left(\frac{\sqrt{1+4 w_j^{(n_0,n_1)}x^2/\zeta_j^2}+2 z^j v_{1,\,j}^{(n_1)} x/\zeta_j+1}{2 \sqrt{1+4 w_j^{(n_0,n_1)}x^2/\zeta_j^2}}\right),
\end{align}
\end{subequations}
where
\begin{subequations}
\label{eq: special mass v w}
\begin{align}
&v_{0,\,j}^{(n_0)} = \prod_{i=1}^{n_0} \mathfrak{p}_j(m_{i+2}), \qquad
v_{1,\,j}^{(n_1)} = \prod_{i=1}^{n_1} \mathfrak{p}_j(m_i),\\
& w_j^{(n_0,n_1)}= v_{0,\,j}^{(n_0)}v_{1,\,j}^{(n_1)}, \qquad \zeta_j=(-1)^{j-1}j!(j-1)!,
\end{align}
\end{subequations}
with the polynomials $\mathfrak{p}_j$ defined as follows:
\begin{equation}
\label{eq: special mass polynomial}
\mathfrak{p}_j(m) = \prod_{k=1-\frac{j}{2}}^{\frac{j}{2}}(m+k).
\end{equation}
These expressions have been tested against the instanton expansion of the combinations appearing in \eqref{eq: What resum} and \eqref{eq: What + W1 resum}, and were found to reproduce the expected pole-contributions to high order. Moreover, their large-$x$ asymptotics are consistent with their role in parametrizing solutions of the Heun equation \eqref{eq: Heun (2,2) zt}. Indeed, approaching a pole of the Coulomb branch parameter $a$ in the resummation ansatz \eqref{eq: What resum} along the real axis, or equivalently taking the limit $x\to \pm\infty$, the functions in \eqref{eq: resumm leading h} behave as
\begin{subequations}
\label{eq: asymptotics w/o W1}
\begin{align}
&h_{1,j}^{+}(\boldsymbol{m},z,x)
=
-\frac{1}{2}\log|x|
+\log\!\left(
\frac{\sqrt{|\zeta_j|}\left(v_{1,\,j}^{(n_1)}\pm (-1)^{j-1}\sqrt{w_j^{(n_0,n_1)}}z^{-j}\right)}
{\sqrt{2}\,v_{1,\,j}^{(n_1)}\,\bigl(w_j^{(n_0,n_1)}\bigr)^{1/4}}
\right)
+\mathcal{O}(|x|^{-1}),\\
&h_{1,j}^{-}(\boldsymbol{m},z,x)
=
\frac{1}{2}\log|x|
+\log\!\left(
\frac{\sqrt{w_j^{(n_0,n_1)}}\mp(-1)^j z^j v_{1,\,j}^{(n_1)}}
{\sqrt{2|\zeta_j|}\,\bigl(w_j^{(n_0,n_1)}\bigr)^{1/4}}
\right)
+\mathcal{O}(|x|^{-1}),
\end{align}
\end{subequations}
while the functions in \eqref{eq: resumm leading ht}, appearing in the resummation ansatz \eqref{eq: What + W1 resum}, behave as
\begin{subequations}
\label{eq: asymptotics w/ W1}
\begin{align}
&\tilde{h}_{1,j}^{+}(\boldsymbol{m},z,x)
=
-\log|x|
+\log\!\left(
\frac{|\zeta_j|\,\Bigl(v_{1,\,j}^{(n_1)}\pm (-1)^{j-1}z^{-j}\sqrt{w_j^{(n_0,n_1)}}\Bigr)}
{2\,v_{1,\,j}^{(n_1)}\sqrt{w_j^{(n_0,n_1)}}}
\right)
+\mathcal{O}(|x|^{-1}),\\[2mm]
&\tilde{h}_{1,j}^{-}(\boldsymbol{m},z,x)
=
\log\!\left(
\frac{1}{2}
\mp\,\frac{(-1)^j z^j v_{1,\,j}^{(n_1)}}{2\sqrt{w_j^{(n_0,n_1)}}}
\right)
+\mathcal{O}(|x|^{-1}) .
\end{align}
\end{subequations}
All other resummation functions $h_{k,j}^{\pm}$ and $\tilde h_{k,j}^{\pm}$, with $k>1$, remain regular in the limits $x\to \pm\infty$, similarly to their bulk analogues. In view of the specific large-$x$ behavior of the resummation functions above, we can define two linearly independent resummed Floquet solutions in slightly different but equivalent ways. These choices differ by $z$-independent normalization factors which may themselves vanish or diverge at $a\to \pm j/2$. This is precisely what is needed to compensate the zeros or divergences produced by a given choice of resummation ansatz. For instance, one natural choice is
\begin{subequations}
\label{eq: resummed heun functions}
\begin{align}
&\tilde{\phi}_0(a,\boldsymbol{m},\hbar;z,\mathfrak{t})=z^{-\frac{a}{\hbar}}\Gamma\br{1+\frac{2a}{\hbar}}\exp\left( \hat{\mathcal{W}}_{0,\, \mathrm{inst}}(a,\boldsymbol{m},\hbar;z,\mathfrak{t}) + \mathcal{W}_{1,\,\mathrm{inst}}(a,\boldsymbol{m},\hbar;\mathfrak{t}) \right),\\
&\tilde{\phi}_1(a,\boldsymbol{m},\hbar;z,\mathfrak{t})=\tilde{\phi}_0(-a,\boldsymbol{m},\hbar;z,\mathfrak{t}),
\end{align}
\end{subequations}
where $\hat{\mathcal{W}}_{0,\,\mathrm{inst}} + \mathcal{W}_{1,\,\mathrm{inst}}$ is given in \eqref{eq: What + W1 resum}, with $\hbar$ reintroduced by an appropriate rescaling of the other variables \eqref{eq:hbar_rescalings}. These solutions are well defined for arbitrary values of the Coulomb branch parameter $a$, up to the ambiguity associated with the branch points at $a\in\frac{1}{2}\mathbb{Z}\setminus \{0\}$. Indeed, in the $a \to j^{\pm}/2$ limit, one finds
\begin{equation}
\label{eq: leading order phi}
\tilde{\phi}_{\pm}^{(j)}(\boldsymbol{m},1;z,\mathfrak{t})=\lim_{a\to j^{\pm}/2}\tilde{\phi}_0(a,\boldsymbol{m},1;z,\mathfrak{t})=\frac{j!}{2}\left(z^{-\frac{j}{2}}\pm\frac{(-1)^j v_{1,\,j}^{(n_1)}}{\sqrt{w_{j}^{(n_0,n_1)}}}z^{\frac{j}{2}}\right)+\mathcal{O}(\mathfrak{t}),
\end{equation}
with the same limit applied to the second solution defined above yielding the same resonant wavefunctions $\tilde{\phi}_{\pm}^{(j)}$, up to an overall normalization. The same conclusion holds upon taking instead the limit $a\to -j^{\pm}/2$, with the roles of $\tilde{\phi}_0$ and $\tilde{\phi}_1$ being exchanged. Thus, for any real value of $a$ away from resonance, the two solutions in \eqref{eq: resummed heun functions} form a basis of linearly independent Floquet solutions, describing eigenfunctions associated with points inside the spectral bands \eqref{eq:dispersions_bands} of the periodic spectral problems related to the Heun equation \eqref{eq: Heun (2,2) zt} and its confluent limits. At a nonzero resonance, each resummed solution admits two lateral limits, giving the resonant (anti)periodic solutions associated with the two corresponding band edges. At $a=0$, by contrast, the two resummed solutions reduce to the same periodic wavefunction corresponding to the minimum of the first spectral band. Since the instanton expansions are regular at $a=0$, this wavefunction can equivalently be obtained by setting $a=0$ before performing the resummation. Generically, its second linearly independent companion is logarithmic and can be constructed by the methods discussed in the next subsection.

One could also consider the solutions that arise naturally by normalizing the defect partition function by the bulk partition function before taking the NS limit:
\begin{subequations}
\label{eq: resummed heun functions psi}
\begin{align}
\psi_0(a,\boldsymbol{m},\hbar;z,\mathfrak{t})
&=
\frac{z^{-\frac{a}{\hbar}}}{\Gamma\left(1-\frac{2a}{\hbar}\right)}
\exp\left(
\hat{\mathcal{W}}_{0,\,\mathrm{inst}}(a,\boldsymbol{m},\hbar;z,\mathfrak{t})
-
\mathcal{W}_{1,\,\mathrm{inst}}(a,\boldsymbol{m},\hbar;\mathfrak{t})
\right),\\
\psi_1(a,\boldsymbol{m},\hbar;z,\mathfrak{t})
&=
\psi_0(-a,\boldsymbol{m},\hbar;z,\mathfrak{t}),
\end{align}
\end{subequations}
where the gamma function at the denominator has been included in order to cancel the singular large-$x$ behavior of the leading resummation functions, which is given by
\begin{subequations}
\begin{align}
&h_{1,j}^{+}(\boldsymbol{m},z,x)-\tilde{f}_{1,j}(\boldsymbol{m},x)
=
\log\!\left(
1\mp (-1)^j\frac{\sqrt{w_j^{(n_0,n_1)}}}{v_{1,\,j}^{(n_1)}}\,z^{-j}
\right)
+\mathcal{O}(|x|^{-1}),\\[2mm]
&h_{1,j}^{-}(\boldsymbol{m},z,x)-\tilde{f}_{1,j}(\boldsymbol{m},x)
=
\log|x|
+\log\!\left(
\frac{\sqrt{w_j^{(n_0,n_1)}}\mp (-1)^j z^j v_{1,\,j}^{(n_1)}}
{|\zeta_j|}
\right)
+\mathcal{O}(|x|^{-1}),
\end{align}
\end{subequations}
as $x\to \pm \infty$.

Before moving to the next subsection, let us note that the factor $w_j^{(n_0,n_1)}$ controls the only square-root branch structure responsible for the splitting of the resummed accessory parameter and Floquet solutions in the resonant limit. For the accessory parameter, this follows from the structure of the bulk resummation functions in \eqref{eq: W0 res fun}. The same square-root factor governs the resummation functions of $\mathcal W_1$ in \eqref{eq: g-tilde functions} and \eqref{eq: f-tilde functions}, as well as the leading defect resummation functions in \eqref{eq: resumm leading h} and \eqref{eq: resumm leading ht}. Together with the linearity of the equations determining the subleading defect resummation functions, this shows that the latter cannot introduce an additional independent square-root branch responsible for the splitting. The vanishing of $w_j^{(n_0,n_1)}$ therefore removes this source of splitting, although it does not by itself guarantee that the resulting resonant limits are regular.

\subsubsection{Logarithmic solutions}
\label{sec:logsol}

At the resonant loci $2a/\hbar = n \in \mathbb Z$, the two $A$-cycle Floquet multipliers coalesce:
\begin{equation}
    \re^{\pm 2\pi i \frac{a}{\hbar}} \to (-1)^{n}.
\end{equation}
Thus, generically, the resonant monodromy is not semisimple and the second independent solution is logarithmic. In the following, we discuss two standard constructions which, when applied to our setting, allow us to obtain the logarithmic companions of the resonant (anti)periodic solutions discussed in the previous subsection.
\paragraph{Reduction of order.}
We begin by recalling the reduction-of-order construction of a logarithmic companion solution. After setting $\hbar=1$ in \eqref{eq: Heun (2,2) zt}, the Heun equation takes the form
\begin{equation}
\psi''+p(z)\psi'+q(z)\psi=0,
\end{equation}
with
\begin{equation}
p(z)=
\frac{1}{z}
-\frac{\mathfrak t}{z}\frac{m_1+m_2}{\mathfrak t-z^{-1}}
+\frac{\mathfrak t}{z}\frac{m_3+m_4}{\mathfrak t-z},
\end{equation}
Thus, by Abel's identity, the Wronskian $W(\psi_1,\psi_2)$ of any two solutions $\psi_1$ and $\psi_2$ is proportional to
\begin{equation}
\mu(z):=\exp\!\left(-\int^z p(\xi)\,d\xi\right)
\propto
z^{-1}(1-\mathfrak t z)^{m_1+m_2}
\left(1-\frac{\mathfrak t}{z}\right)^{m_3+m_4},
\end{equation}
where the proportionality constant is independent of $z$. Let $\psi^{(n)}$ be a nonzero resonant Floquet solution,
\begin{equation}
\psi^{(n)}(z\re^{2\pi i})
=
(-1)^n\psi^{(n)}(z).
\end{equation}
Once the normalization of $\mu(z)$ has been fixed, we define its canonical companion by
\begin{equation}
\psi_{\log}^{(n)}(z)
:=
\psi^{(n)}(z)
\int^z
\frac{\mu(w)}{\psi^{(n)}(w)^2}\,dw, 
\end{equation}
which, by construction, satisfies
\begin{equation}
W\!\left(\psi^{(n)},\psi_{\log}^{(n)}\right)=\mu(z).
\end{equation}
Therefore, for fixed $\psi^{(n)}$, the companion is uniquely determined by this Wronskian normalization up to
\begin{equation}
\psi_{\log}^{(n)}
\longmapsto
\psi_{\log}^{(n)}+c(\boldsymbol{m}, \mathfrak{t})\,\psi^{(n)},
\end{equation}
where this ambiguity corresponds to changing the lower endpoint of the integral. In the annulus $|\mathfrak t|<|z|<|\mathfrak t|^{-1}$, the integrand admits a Laurent expansion in $z$ and, upon integration, the coefficient of $z^{-1}$ is the only term that produces a logarithm. Thus one can write
\begin{equation}
\psi_{\log}^{(n)}(z)
=
K^{(n)}(\boldsymbol m,\mathfrak t)\,
\psi^{(n)}(z)\log z
+
\varphi^{(n)}(z),
\end{equation}
where $\varphi^{(n)}$ has the same $A$-cycle monodromy as $\psi^{(n)}$,
\begin{equation}
\varphi^{(n)}(ze^{2\pi i})
=
(-1)^n \varphi^{(n)}(z),
\end{equation}
and
\begin{equation}
\label{eq: K coeff}
K^{(n)}(\boldsymbol m,\mathfrak t)
= \frac{1}{2\pi i}\oint_{\gamma} \frac{\mu(z)}
{\psi^{(n)}(z)^2}\,dz,
\end{equation}
where $\gamma$ is a positively oriented closed contour contained in the annulus $|\mathfrak t|<|z|<|\mathfrak t|^{-1}$ and winding once around the origin $z=0$. It follows that the canonical logarithmic solution transforms under the $A$-cycle monodromy as
\begin{equation}
\psi_{\log}^{(n)}(z\re^{2\pi i})
=
(-1)^n\psi_{\log}^{(n)}(z)
+
2\pi i\,(-1)^n K^{(n)}(\boldsymbol m,\mathfrak t)\,
\psi^{(n)}(z),
\end{equation}
and, in the basis
\begin{equation}
\bigl(\psi^{(n)},\psi_{\log}^{(n)}\bigr)^{\mathrm{T}},
\end{equation}
one obtains
\begin{equation}
M_A
=
(-1)^n
\begin{pmatrix}
1 & 0 \\
2\pi i\,K^{(n)} & 1
\end{pmatrix}.
\end{equation}
Thus, the coefficient $K^{(n)}$ measures the nilpotent part of the resonant $A$-cycle monodromy. In particular,
\begin{equation}
\label{eq:Kvanishing}
K^{(n)}=0
\qquad\Longleftrightarrow\qquad
M_A=(-1)^n I,
\end{equation}
provided that, at the zeros of $K^{(n)}$, the accessory parameter and both solutions $\psi^{(n)}$ and $\psi_{\log}^{(n)}$ remain regular, and that their limiting values continue to form a linearly independent pair. In that case, the resonant monodromy is semisimple, or equivalently there exists a second independent resonant Floquet solution with the same multiplier.

However, as we discuss in more detail in the next subsection, $K^{(n)}$ is not an invariant scalar until a normalization of the resonant Floquet solution has been fixed. Indeed, keeping the normalization of $\mu(z)$ fixed, a parameter-dependent rescaling
\begin{equation}
\psi^{(n)}
\longmapsto
f(\boldsymbol m,\mathfrak t)\,\psi^{(n)}
\end{equation}
induces
\begin{equation}
K^{(n)}
\longmapsto
f(\boldsymbol m,\mathfrak t)^{-2}K^{(n)}.
\end{equation}
Thus $K^{(n)}$ should be regarded as the local representative, in a chosen local frame for the resonant Floquet line, of the nilpotent monodromy coefficient. In particular, its local expression depends on the frame, while its vanishing is invariant under invertible changes of frame.

Applying this construction to our setting, for any of the resonant representatives constructed above, such as $\psi^{(n)} = \tilde{\phi}_{\pm}^{(n)}$, we define\footnote{We recall that the sign $\pm$ in the wavefunctions $\tilde{\phi}_{\pm}^{(n)}$ labels solutions of two distinct spectral problems, since for generic values of the mass parameters the corresponding accessory parameters $u_{\pm}^{(n)}$ are different.}
\begin{equation}
\label{eq: can_log_sol_pm}
\tilde{\phi}_{\pm,\,\log}^{(n)}(z)
:=
\tilde{\phi}_{\pm}^{(n)}(z)
\int^z
\frac{\mu(w)}{\tilde{\phi}_{\pm}^{(n)}(w)^2}\,dw
=
K_{\pm}^{(n)}(\boldsymbol m,\mathfrak t)\,
\tilde{\phi}_{\pm}^{(n)}(z)\log z
+
\tilde{\varphi}_{\pm}^{(n)}(z),
\end{equation}
where $K_{\pm}^{(n)}$ is obtained from \eqref{eq: K coeff} by replacing
$\psi^{(n)}$ with $\tilde{\phi}_{\pm}^{(n)}$. The gauge-theory construction of the previous subsection gives the resonant solution, and hence its logarithmic companion together with the coefficient $K_{\pm}^{(n)}$, as an expansion in powers of $\mathfrak t$. However, it does not by itself specify a preferred global normalization of the resonant solution. For this reason, in practice one might have to impose a different normalization a posteriori, order by order in $\mathfrak t$, for example by fixing a chosen Laurent coefficient of $\tilde{\phi}_{\pm}^{(n)}$. Once a choice has been made, one can compute the corresponding representative of $K_{\pm}^{(n)}$ and the associated logarithmic solution as power series in $\mathfrak t$.

Different choices of normalization will lead to different local  representatives. In particular, if a zero of the intrinsic nilpotent coefficient is not visible in a given representative, this may indicate that the chosen frame is singular at that locus: the rescaling used to define the normalized resonant solution may have introduced an artificial pole or zero. One should then pass to a different local normalization, regular on the locus under consideration, and recompute the corresponding representative of $K_{\pm}^{(n)}$.

\paragraph{Coalescence.}

In the generic non-semisimple case, an alternative construction of the logarithmic solution is obtained by coalescing two non-resonant independent Floquet solutions. If two limiting solutions $\phi_0$ and $\phi_1$ become proportional at resonance, one may choose a $z$-independent factor $A(a,\boldsymbol m;\mathfrak t)$ such that
\begin{equation}
\lim_{2a\to n^\pm}
\left(
\phi_0(a;z)-A(a,\boldsymbol m;\mathfrak t)\phi_1(a;z)
\right)=0.
\end{equation}
Then the standard coalescence construction gives a second solution of the resonant equation:
\begin{equation}
\phi_{\pm,\, \log}^{(n)}(z)
:=
\lim_{2a\to n^\pm}
\frac{
\phi_0(a;z)-A(a,\boldsymbol m;\mathfrak t)\phi_1(a;z)
}{2a-n},
\end{equation}
whenever the limit exists. Equivalently, if the numerator extends holomorphically to the resonant point, this may be written as
\begin{equation}
\label{eq:collision_deriv_diff}
\phi_{\pm,\, \log}^{(n)}(z)
:=
\lim_{2a\to n^\pm}\frac{1}{2}\partial_a \br{\phi_0(a;z)-A(a,\boldsymbol m;\mathfrak t)\phi_1(a;z)}.
\end{equation}
This is the usual Frobenius/Floquet mechanism by which a logarithmic solution appears when two exponents collide. In the present problem, however, it cannot be applied directly before resummation, because the $\mathfrak t$-expanded Floquet solutions develop poles of increasing order at the resonant values of $a$. Instead, after resummation, the numerator solves the Heun equation \eqref{eq: Heun (2,2) zt} for every non-resonant value of $a$, and its divided limit solves the limiting equation at resonance. The monodromy of $\phi_{\pm,\, \log}^{(n)}$ is fixed solely by differentiating the two Floquet multipliers. If
\begin{equation}
\phi_{\pm}^{(n)}(z)
=
\lim_{2a\to n^\pm}\phi_0(a;z)
=
\lim_{2a\to n^\pm}A(a,\boldsymbol m;\mathfrak t)\phi_1(a;z),
\end{equation}
then one finds, up to the overall sign convention in the definition of $\phi_{\pm,\, \log}^{(n)}$ and in the assignment of the two Floquet multipliers,
\begin{equation}
\phi_{\pm,\, \log}^{(n)}(z\re^{2\pi i})
=
(-1)^n\phi_{\pm,\, \log}^{(n)}(z)
-2\pi i(-1)^n\phi_{\pm}^{(n)}(z).
\end{equation}
Thus, the logarithmic companion obtained in this collision basis has a mass-independent representative for the off-diagonal monodromy coefficient, due to the normalization of the logarithmic pair selected by the collision construction.

\paragraph{Single-derivative construction.}

In our setting, we can consider a slight variation of the previous construction, suggested by the analytic properties of the resummed accessory parameter $u^{(n_0,n_1)}_{\mathrm{res}}$. Instead of differentiating a vanishing difference of two colliding solutions, one may try to differentiate a single Floquet solution. In general, this does not give another solution, since the Heun operator in \eqref{eq: Heun (2,2) zt} does not commute with $\partial_a$. However, the dependence on $a$ enters the differential operator only through the accessory parameter $u^{(n_0,n_1)}_{\mathrm{res}}$. Hence, after differentiating the equation with respect to $a$, the source term is proportional to $\partial_a u^{(n_0,n_1)}_{\mathrm{res}}$ and if the resummed accessory parameter satisfies
\begin{equation}
\label{eq: vanishing du/da}
\lim_{2a\to n}\partial_a u^{(n_0,n_1)}_{\mathrm{res}} =0,
\end{equation}
then the inhomogeneous term vanishes in the resonant limit. Thus, if the prefactor $B(a,\boldsymbol m;\mathfrak t)$ is chosen so
that both $B\phi_0$ and $B\partial_a\phi_0$ admit finite resonant limits, the function
\begin{equation}
\label{eq: da log def}
\chi_{\pm,\,\mathrm{log}}^{(n)}(z)
:=
\lim_{2a\to n^\pm}
B(a,\boldsymbol m;\mathfrak t)\,
\partial_a\phi_0(a;z)
\end{equation}
is again a solution of the resonant Heun equation.

Its logarithmic monodromy follows again from differentiating the Floquet multiplier. More precisely, denoting the resonant solution selected by the same prefactor by
\begin{equation}
\chi_{\pm}^{(n)}(z)
:=
\lim_{2a\to n^\pm}
B(a,\boldsymbol m;\mathfrak t)\,
\phi_0(a;z),
\end{equation}
then, up to the overall sign convention for the Floquet multiplier of $\phi_0$, one obtains
\begin{equation}
\chi_{\pm,\,\mathrm{log}}^{(n)}(z\re^{2\pi i})
=
(-1)^n\chi_{\pm,\,\mathrm{log}}^{(n)}(z)
-2\pi i (-1)^n\chi_{\pm}^{(n)}(z).
\end{equation}
As usual, $\chi_{\pm,\,\mathrm{log}}^{(n)}$ is defined only up to the addition of a multiple of the resonant solution, and this ambiguity depends on the chosen prefactor $B$. In this sense, the single-derivative construction is a stronger statement than the standard collision construction: it uses the special property \eqref{eq: vanishing du/da} of the resummed accessory parameter. The condition \eqref{eq: vanishing du/da} follows from a simple conjectural property of the resummed dual quantum period. Let
\begin{equation}
a_D^{(n_0,n_1)}=\partial_a\mathcal{W}_0^{(n_0,n_1)} = a_{D,\,\mathrm{pert}}^{(n_0,n_1)}+a_{D,\,\mathrm{inst}}^{(n_0,n_1)}
\end{equation}
denote the dual period obtained from the resummed twisted superpotential, including the instanton contribution $a_{D,\,\mathrm{inst}}^{(n_0,n_1)}$ obtained from the $a$-derivative of \eqref{eq: W0 resum}, together with the classical and one-loop terms:
\begin{equation}
\label{eq: aD_pert}
    a_{D,\,\mathrm{pert}}^{(n_0,n_1)}=-4 a \log \mathfrak{t}+2\log \frac{\Gamma(1+2a)}{\Gamma(1-2a)}-\sum_{i_0=1}^{n_0}\log \frac{\Gamma(1+m_{i_0+2}+a)}{\Gamma(1+m_{i_0+2}-a)}-\sum_{i_1=1}^{n_1}\log \frac{\Gamma(1+m_{i_1}+a)}{\Gamma(1+m_{i_1}-a)}.
\end{equation}
We conjecture that, at every resonant point,
\begin{equation}
\label{eq:conj_ad}
    \lim_{a\to \frac{n}{2}}\exp\,\left(a_D^{(n_0,n_1)}\right)=1,
    \qquad n\in\mathbb{Z}.
\end{equation}
Since the limiting value of $a_D^{(n_0,n_1)}$ is independent of $\mathfrak t$, this implies that
\begin{equation}
\lim_{a\to n/2}\mathfrak t\,\partial_{\mathfrak t}a_D^{(n_0,n_1)}=0,
\end{equation}
and therefore \eqref{eq: vanishing du/da}. We tested this conjecture in the $N_f = (2,2)$, $N_f = (1,1)$ and $N_f = (0,0)$ theories by computing the resummed instanton NS prepotential to the highest accessible order in the resummation functions in each case. Moreover, the explicit form of the leading resummation functions in \eqref{eq: W0 res fun} is consistent with this conjecture. In particular, these functions cancel the singularities arising from the Gamma-function contributions in \eqref{eq: aD_pert} at the resonant points.

\subsection{Special mass loci and semisimple resonant monodromy}
\label{sec: special mass loci}

Let us now consider the loci in the $\boldsymbol m$-parameter space on which the resonant $A$-cycle monodromy may become semisimple, working at the level of series expansions around $\mathfrak t=0$, and setting again $\hbar=1$ in order to simplify the notation. The $\hbar$-dependence can be restored by rescaling all parameters appropriately. We consider a nonzero resonant value $2a=n$, with $n\in\mathbb{Z}\setminus\{0\}$, and write $N:=|n|$. The case $n=0$ will not be discussed here since, as mentioned before, the instanton functions considered below are regular at $a=0$, so the resonant limit can already be taken term by term in the instanton expansion.

As already discussed above, for a fixed resonant branch $\sigma\in\{+,-\}$, the coefficient $K_{\sigma}^{(n)}$ is a local representative of the nilpotent part of the resonant $A$-cycle monodromy. Its value depends on the chosen normalization of the resonant Floquet solution, but its vanishing is invariant under regular changes of local frame. Thus, the intrinsic semisimplicity condition is
\begin{equation}
    K_{\sigma}^{(n)}=0,
\end{equation}
together with the requirement that, at the zeros of $K_{\sigma}^{(n)}$, the corresponding resonant accessory parameter, as well as the (anti-)periodic and logarithmic solutions, remain regular and that their limits continue to form a basis of solutions.

The problem of determining this locus can already be approached at the level of the instanton expansions of the NS prepotentials $\mathcal{W}_0$, $\mathcal{W}_1$, and $\hat{\mathcal{W}}_\beta$. In what follows, all statements refer to the standard small-$\mathfrak{t}$ instanton expansion and have been checked only to finite order. We start from the twisted effective superpotential $\mathcal{W}_0$, which determines the accessory parameter through \eqref{eq: qmatone}. Since this NS function is even in $a$, it is enough to study the positive resonant limit $2a\to N$. At all orders checked, the coefficients of the highest-order pole contributions at $2a=N$ are proportional to the mass factor
\begin{equation}
\label{eq: special mass factor w}
w_{N}^{(n_0,n_1)}
=
\prod_{i=1}^{n_0} \mathfrak{p}_{N}(m_{i+2})
\prod_{j=1}^{n_1} \mathfrak{p}_{N}(m_j),
\end{equation}
where the polynomial $\mathfrak{p}_{N}$ is defined in \eqref{eq: special mass polynomial}. Thus the vanishing of one factor in $w_{N}^{(n_0,n_1)}$ removes the leading source of singularities but it does not in general make the instanton expansion regular in the limit $a\to N/2$.

For example, in the $N_f=(2,2)$ theory\footnote{We recall that the bulk instanton prepotentials $\mathcal W_0$ and $\mathcal W_1$ are symmetric under permutations of the mass parameters $\boldsymbol m=(m_1,m_2,m_3,m_4)$, whereas the functions $\widehat{\mathcal W}_\beta$ are symmetric only under separate permutations of the pairs $(m_1,m_2)$ and $(m_3,m_4)$.}, if $m_1$ is specialized so that $\mathfrak{p}_{N}(m_1)=0$, the residues of the remaining poles at $a=N/2$ are found to contain the common factor
\begin{equation}
\label{eq:p-factor}
   \widetilde{w}_{N}^{(2,2)}=\mathfrak{p}_{N}(m_2)\mathfrak{p}_{N}(m_3)\mathfrak{p}_{N}(m_4).
\end{equation}
Thus, a further specialization of one of the remaining mass parameters completely removes the remaining poles at $a = N/2$ in $\mathcal{W}_{0,\,\mathrm{inst}}$. An analogous pole-cancellation pattern holds for $\mathcal{W}_{1,\,\mathrm{inst}}$ in the $N_f=(2,2)$ theory and, by taking appropriate decoupling limits, for the bulk NS functions of theories with fewer hypermultiplets. 

Let $I_{n_0,n_1}:= \{1,\ldots,n_1\}\cup\{3,\ldots,n_0+2\}$ denote the set of labels of the hypermultiplet masses present in the $N_f=(n_0,n_1)$ theory. We find that, on the locus
\begin{equation}
\label{eq: bulk special locus}
\mathcal D_{\mathrm{bulk}}^{(N)}
:=
\bigcup_{\substack{i,j\in I_{n_0,n_1}\\ i<j}}
\left\{
\boldsymbol m\ \middle|\ 
\mathfrak p_N(m_i)=0,\quad
\mathfrak p_N(m_j)=0
\right\},
\end{equation}
the small-$\mathfrak t$ expansions of both bulk instanton prepotentials $\mathcal W_{0,\,\mathrm{inst}}$ and $\mathcal W_{1,\,\mathrm{inst}}$ admit a regular resonant limit at $2a=n$. Consequently, the same holds for the accessory parameter $u^{(n_0,n_1)}$. Complete cancellation of the bulk poles therefore requires the specialization of at least two distinct mass parameters. In particular
\begin{equation}
\mathcal D_{\mathrm{bulk}}^{(N)}=\varnothing \quad \text{if} \quad n_0+n_1\leq 1.
\end{equation}

At the level of the small-$\mathfrak t$ instanton expansion, $\mathcal D_{\mathrm{bulk}}^{(N)}$ is precisely the locus on which the $N$-th spectral gap closes. Indeed, for $\boldsymbol m_\ast\in\mathcal D_{\mathrm{bulk}}^{(N)}$, the two lateral accessory parameters coincide order by order in $\mathfrak t$,
\begin{equation}
\label{eq:eq_at_gap_clos}
u^{(n)}(\boldsymbol m_\ast;\mathfrak t)
:=
u_+^{(n)}(\boldsymbol m_\ast;\mathfrak t)
=
u_-^{(n)}(\boldsymbol m_\ast;\mathfrak t),
\qquad N=|n|.
\end{equation}
Furthermore, these gap-closing loci are nested separately for odd and even values of $N$. Indeed, $\mathfrak p_N(m)$ divides $\mathfrak p_{N+2r}(m)$ for $r\in\mathbb Z_{\geq0}$, and therefore
\begin{equation}
\mathcal D_{\mathrm{bulk}}^{(N)}
\subseteq
\mathcal D_{\mathrm{bulk}}^{(N+2r)},
\qquad r\in\mathbb Z_{\geq0}.
\end{equation}
Thus, a mass specialization belonging to $\mathcal D_{\mathrm{bulk}}^{(N)}$ closes all gaps indexed by $N+2r$, $r\in\mathbb Z_{\geq0}$. This does not in
general define a finite-gap potential, since the gaps of the opposite parity may remain open.

We now turn to the instanton defect prepotentials $\hat{\mathcal{W}}_{\beta,\, \mathrm{inst}}$. Since the two choices of defect NS functions are related by $a\to -a$, it is enough to discuss $\beta=0$. In this case, the coefficients of the leading poles at the negative and positive resonances, $2a=-N$ and $2a=N$, are proportional, respectively, to 
\begin{equation}
    v_{0,\,N}^{(n_0)}
    =
    \prod_{i=1}^{n_0}\mathfrak{p}_{N}(m_{i+2}),
    \qquad
    v_{1,\,N}^{(n_1)}
    =
    \prod_{i=1}^{n_1}\mathfrak{p}_{N}(m_i),
\end{equation}
as defined in \eqref{eq: special mass v w}. For $\beta=1$ the roles of these two factors are exchanged, as a consequence of the relation $a\to -a$. Once one mass has been tuned to cancel the leading-order poles, for instance at the negative resonance $2a=-N$, the residues of all remaining poles at the same resonance vanish upon specializing any additional mass parameter to a zero of $\mathfrak{p}_{N}(m)$. The same holds for positive resonances.

Not every component of the bulk locus $\mathcal D_{\mathrm{bulk}}^{(N)}$ is therefore compatible with the regularity of the defect prepotentials, at least at the level of their instanton expansion. By requiring $\mathcal{W}_{0,\, \mathrm{inst}}$, $\mathcal{W}_{1,\, \mathrm{inst}}$, and both defect prepotentials $\hat{\mathcal{W}}_{\beta,\, \mathrm{inst}}$ to have a finite limit as $a\to n/2$, the instanton expansion singles out the smaller locus
\begin{equation}
\label{eq: defect special locus}
    \mathcal D_{\mathrm{defect}}^{(N)}
    :=
    \bigcup_{i=1}^{n_1}\bigcup_{j=1}^{n_0}
    \left\{\boldsymbol{m}\ \middle|\
    \mathfrak{p}_{N}(m_i)=0,\; \mathfrak{p}_{N}(m_{j+2})=0
    \right\} \subseteq \mathcal D_{\mathrm{bulk}}^{(N)}.
\end{equation}
Equivalently, one must require the vanishing of at least one factor in $v_{1,\,N}^{(n_1)}$ and at least one factor in $v_{0,\,N}^{(n_0)}$. In particular
\begin{equation}
\mathcal D_{\mathrm{defect}}^{(N)}=\varnothing \quad \text{if} \quad n_0=0 \text{ or } n_1=0,
\end{equation}
and, similarly to the bulk case,
\begin{equation}
\mathcal D_{\mathrm{defect}}^{(N)}
\subseteq
\mathcal D_{\mathrm{defect}}^{(N+2r)},
\qquad r\in\mathbb Z_{\geq0}.
\end{equation}
For example, in the $N_f=(2,2)$ theory this means that one mass from the pair $(m_1,m_2)$ and one mass from the pair $(m_3,m_4)$ must be specialized to zeros of $\mathfrak{p}_{N}(m)$. On this locus, the pole cancellations of the bulk and defect instanton
prepotentials give two regular and linearly independent resonant
limits, one obtained from each defect function. Hence $\mathcal D_{\mathrm{defect}}^{(N)}$ is contained in the semisimple locus of the resonant $A$-cycle monodromy.

Conversely, since semisimplicity requires the corresponding spectral gap to close, any additional component of the semisimple locus detected by the small-$\mathfrak t$ expansion would have to lie in $\mathcal D_{\mathrm{bulk}}^{(N)}\setminus\mathcal D_{\mathrm{defect}}^{(N)}$. For example, suppose that the masses are chosen so that the poles of $\hat{\mathcal W}_0$ at the negative resonance $2a=-N$ are cancelled. One then obtains the regular accessory parameter $u^{(n)}(\boldsymbol m;\mathfrak t)$ in \eqref{eq:eq_at_gap_clos}, together with a resonant (anti)periodic solution from the corresponding limit of the wavefunction constructed from $\hat{\mathcal W}_0$. The wavefunction constructed from $\hat{\mathcal W}_1$ instead diverges in the same resonant limit. Multiplying it by a suitable $z$-independent normalization factor, which vanishes as $2a\to-N$, produces a finite limit proportional to the solution obtained from $\hat{\mathcal W}_0$. It therefore does not provide a second linearly independent resonant Floquet solution and, within the small-$\mathfrak t$ expansion, the semisimple locus coincides with $\mathcal D_{\mathrm{defect}}^{(N)}$. We will provide further evidence for this statement below.

From the point of view of the instanton expansion, the natural order of operations is to first specialize the masses to a point $\boldsymbol m_\ast\in\mathcal D_{\mathrm{defect}}^{(N)}$ and only then take the resonant limit $2a\to n$. The resummation procedure gives access to the complementary order of limits. At generic masses one first resums the singular instanton series and then takes the one-sided limits $2a\to n^\pm$. This produces, for example, the two resonant solutions $\tilde\phi_{\pm}^{(n)}$ in \eqref{eq: leading order phi}, which are in general associated with two different limiting accessory parameters $u_{\pm}^{(n)}$ and therefore with two different resonant Heun equations. However, it is already clear from explicit examples of resummation functions, such as $g_{2,1}$ in \eqref{eq:g21_(1,1)} and $\tilde{h}_{2,1}^{\pm}$ in \eqref{eq:example_resum_(1,1)}, that, in general, the resonant limit does not commute with the limit $\boldsymbol{m}\to\boldsymbol{m}_\ast\in\mathcal D^{(N)}_{\mathrm{defect}}$. Nevertheless, since each resummation function captures infinitely many terms in the instanton expansion, they still provide non-trivial information on the semisimple mass loci of interest.

Indeed, the leading resummation functions show the same mass factors that appeared in the previous instanton analysis. For the bulk twisted superpotential $\mathcal{W}_0$, the leading function $g_{1,N}$ depends on $w_{N}^{(n_0,n_1)}$ which is precisely the factor controlling the leading bulk poles. Similarly, the leading singular part of $\mathcal W_{1,\,\mathrm{inst}}$ is governed by $\tilde f_{1,N}$ and contains the same factor $w_{N}^{(n_0,n_1)}$. For the defect prepotentials, the leading functions $h_{1,N}^{\pm}$ and $\tilde h_{1,N}^{\pm}$ distinguish the two factors $v_{0,\,N}^{(n_0)}$ and $v_{1,\,N}^{(n_1)}$, in agreement with the fact that the two defect functions distinguish between positive and negative resonant values of $a$. More precisely, when $v_{0,N}^{(n_0)}=0$, the functions $h_{1,N}^{+}$ and $\tilde h_{1,N}^{+}$ vanish, whereas 
\begin{equation}
\label{eq:leading_sing_minus}
h_{1,N}^{-} = \tilde h_{1,N}^{-} = \log\br{1+\frac{z^N v_{1,\,N}^{(n_1)}}{\zeta_N}x} .
\end{equation}
Conversely, when $v_{1,N}^{(n_1)}=0$, the functions $h_{1,N}^{-}$ and $\widetilde h_{1,N}^{-}$ vanish, while
\begin{equation}
\label{eq:leading_sing_plus}
h_{1,N}^{+} = \tilde h_{1,N}^{+} = \log\br{1+\frac{ v_{0,\,N}^{(n_0)}}{z^N \zeta_N}x} .
\end{equation}

As discussed in Section~\ref{sec: leading order}, $w_{N}^{(n_0,n_1)}$ is the only factor controlling the square-root branch cuts responsible for the splitting of the resummed accessory parameter and wavefunctions in the resonant limit. When $w_{N}^{(n_0,n_1)}=0$, the square-root branch cuts disappear from the bulk and defect resummation functions, although these functions may still fail to admit a finite resonant limit if all poles in the instanton expansion have not been cancelled. This is manifest in the examples given in \eqref{eq:leading_sing_minus} and \eqref{eq:leading_sing_plus}. 

Thus, the leading resummation functions are consistent with the instanton-level pole-cancellation pattern described above. On the locus $\mathcal D_{\mathrm{defect}}^{(N)}$, where at least one factor in $v_{0,\,N}^{(n_0)}$ and one factor in $v_{1,\,N}^{(n_1)}$ vanish, the branch-point singularities of the resummed NS functions are expected to disappear. Since $\mathcal D_{\mathrm{defect}}^{(N)} \subseteq\mathcal D_{\mathrm{bulk}}^{(N)}$, the accessory parameters already coincide as in \eqref{eq:eq_at_gap_clos}. Correspondingly, by first specializing the mass parameters and only then taking the resonant limit in the two resummed defect solutions, one obtains two resonant wavefunctions that are expected to give linearly independent (anti)periodic solutions of the same resonant Heun equation.

Let us instead consider the case in which the mass parameters $\boldsymbol m$ belong to $\mathcal D_{\mathrm{bulk}}^{(N)} \setminus\mathcal D_{\mathrm{defect}}^{(N)}$. Then $w_N^{(n_0,n_1)}=0$, while exactly one of the two factors $v_{0,\,N}^{(n_0)}$ and $v_{1,\,N}^{(n_1)}$ vanishes. Let us assume that $v_{1,\,N}^{(n_1)}=0$, the other case being analogous. The resummed accessory parameter then admits a finite and unambiguous resonant limit as $2a\to N$, and the same holds for one of the resummed Floquet solutions, say $\tilde\phi_0$ in \eqref{eq: resummed heun functions}. By contrast, the corresponding limit of the second resummed solution $\tilde\phi_1$ diverges, as is clear from \eqref{eq:leading_sing_plus}. Since the accessory parameter remains regular, this divergence reflects a singular choice of normalization of $\tilde\phi_1$. It must therefore be removable, at the level of the small-$\mathfrak t$ expansion, by multiplying $\tilde\phi_1$ by a suitable $z$-independent normalization factor. The leading behavior in \eqref{eq:leading_sing_plus} then shows that the renormalized limit is not linearly independent of the limit of $\tilde\phi_0$. If the same proportionality persists order by order after including the subleading resummation functions, the two resummed Floquet solutions yield only one resonant (anti)periodic solution, with the second linearly independent solution being logarithmic. Then, within the small-$\mathfrak t$ expansion, the semisimple locus coincides with $\mathcal D_{\mathrm{defect}}^{(N)}$.

\subsection{Example: \texorpdfstring{$N_f = (1,1)$}{Nf=(1,1)}}

In Appendix \ref{app: example} we illustrate the resummation procedure explicitly in the $N_f=(1,1)$ theory, deriving the resummation functions $\tilde h$ appearing in \eqref{eq: What + W1 resum}. The relation \eqref{eq: relation h-funs} then gives the corresponding $h$-functions entering \eqref{eq: What resum}, as shown at the end of Appendix \ref{app: example}. In this subsection we use these results to display the first few terms in the $\mathfrak{t}$-expansion of certain resonant solutions together with their logarithmic companions. We begin with the $N_f=(1,1)$ Heun equation,
\begin{equation}
\label{eq: Heun (1,1) zt}
\left[
\partial_z^2
+
\left(
\mathfrak{t}
+\frac{1}{z}
-\frac{\mathfrak{t}}{z^2}
\right)\partial_z
-
\left(
\frac{m_1\mathfrak{t}}{z}
-\frac{u^{(1,1)}}{z^2}
+\frac{m_3\mathfrak{t}}{z^3}
\right)
\right]\psi(z)=0, \qquad \hbar = 1,
\end{equation}
where the accessory parameter is determined by the twisted effective superpotential as
\begin{equation}
u^{(1,1)}= 
\frac{1}{2}\mathfrak{t}\,\partial_\mathfrak{t} \mathcal{W}_{0}^{(1,1)}.
\end{equation}
Using the resummed expression \eqref{eq: W0 resum} for the instanton part of $\mathcal{W}_{0}^{(1,1)}$, we obtain the resonant accessory parameters by
\begin{equation}
    u_{\pm}^{(n)} = \lim_{2a\to n^{\pm}}u^{(1,1)}_{\mathrm{res}}.
\end{equation}
It is enough to consider positive resonances, $n>0$. Indeed, since the accessory parameter is an even function of $a$, the negative resonances are related to the positive ones by
\begin{equation}
\lim_{2a\to n^{\pm}}u^{(1,1)}_{\mathrm{res}}
=
\lim_{2a\to (-n)^{\mp}}u^{(1,1)}_{\mathrm{res}}.
\end{equation}
For the first two resonances this gives
\begin{subequations}
\begin{align}
 u_{\pm}^{(1)}
&=
\mp \frac{1}{2}
\sqrt{(2m_1+1)(2m_3+1)}\,\mathfrak{t}
+
\frac{1}{8}
\left(4m_1m_3+2m_1+2m_3-3\right)\mathfrak{t}^2
+
\mathcal{O}(\mathfrak{t}^3),
\\
u_{\pm}^{(2)}
&=
-\frac{1}{3}
\left(
2+m_1+m_3+2m_1m_3
\pm 3\sqrt{m_1(m_1+1)m_3(m_3+1)}
\right)\mathfrak{t}^2
+
\mathcal{O}(\mathfrak{t}^4).
\end{align}
\end{subequations}
Here, and in the examples below, we take $m_1,m_3\geq 0$ for simplicity. The corresponding resonant solutions can be obtained, for example, by taking the same limits of $\tilde{\phi}_0$ in \eqref{eq: resummed heun functions}. For the first two positive 
resonances one finds
\begin{equation}
\begin{aligned}
\tilde{\phi}_{\pm}^{(1)}
&= \frac{1}{2}\br{
z^{-\frac{1}{2}}
\mp 
\sqrt{\frac{1+2m_1}{1+2m_3}}\,z^{\frac{1}{2}}
}
+ \frac{1}{4}\left[
-\frac{1-2m_3}{2}\,z^{-\frac{3}{2}}
\mp 
\frac{1-4m_1m_3}{\sqrt{(1+2m_1)(1+2m_3)}}\,z^{-\frac{1}{2}}
\right. \\
&\qquad\left.
+
\frac{(1+2m_1)(1-2m_3)}{1+2m_3}\,z^{\frac{1}{2}}
\pm
\frac{(1-2m_1)(1+2m_1)}{2\sqrt{(1+2m_1)(1+2m_3)}}\,z^{\frac{3}{2}}
\right] \mathfrak{t}
+ \mathcal{O}(\mathfrak{t}^2),
\end{aligned}
\end{equation}
\begin{equation}
\begin{aligned}
\tilde{\phi}_{\pm}^{(2)}
&=
z^{-1}
\pm
\sqrt{\frac{m_1(m_1+1)}{m_3(m_3+1)}}\,z
+
\left[
\frac{m_3-1}{3}\,z^{-2}
-1-m_1
\right. \\
&\qquad\left.
\mp
\sqrt{\frac{m_1(m_1+1)(m_3+1)}{m_3}}
\pm
\frac{m_1(m_1^2-1)}
{3\sqrt{m_1(m_1+1)m_3(m_3+1)}}\,z^2
\right]\mathfrak{t}
+\mathcal{O}(\mathfrak{t}^2).
\end{aligned}
\end{equation}
Equivalently, one may take the same resonant limits at the level of $\tilde{\phi}_1$ in \eqref{eq: resummed heun functions}. The resulting solutions agree with those above up to an overall normalization independent of $\mathfrak{t}$, at least to the order displayed.

Following the discussion of the previous sections, the resonant solutions above determine the $\mathfrak{t}$-expansion of their canonical logarithmic companions. Using \eqref{eq: can_log_sol_pm}, we find
\begin{equation}
\begin{aligned}
\tilde{\phi}_{\pm,\,\mathrm{log}}^{(1)}
&=
\pm\,2\sqrt{\frac{1+2m_3}{1+2m_1}}\,z^{-\frac{1}{2}}
+\left[
\pm\frac{4m_3^2-1}
{2\sqrt{(1+2m_1)(1+2m_3)}}\,z^{-\frac{3}{2}}
\right.\\
&\qquad
-\frac{28m_1m_3+6m_1+10m_3+1}
{2(1+2m_1)}\,z^{-\frac{1}{2}}
\pm\frac{2m_1-1}{2}
\sqrt{\frac{1+2m_3}{1+2m_1}}\,z^{\frac{1}{2}}
\\
&\qquad\left.
+
\left(
-(1+2m_3)\,z^{-\frac{1}{2}}
\pm
\sqrt{(1+2m_1)(1+2m_3)}\,z^{\frac{1}{2}}
\right)\log z
\right]\mathfrak{t}
+\mathcal{O}(\mathfrak{t}^2),
\end{aligned}
\end{equation}
and
\begin{equation}
\begin{aligned}
\tilde{\phi}_{\pm,\,\mathrm{log}}^{(2)}
&=
\mp\frac{1}{2}
\sqrt{\frac{m_3(1+m_3)}{m_1(1+m_1)}}\,z^{-1}
\\
&\quad
\mp\frac{1}{6}
\sqrt{\frac{m_3(1+m_3)}{m_1(1+m_1)}}
\left[
(m_3-1)\,z^{-2}
-3(1+m_1)
\right]\mathfrak{t}
+\mathcal{O}(\mathfrak{t}^2).
\end{aligned}
\end{equation}
For the second resonance the logarithmic term first appears at order $\mathfrak{t}^2$, and therefore requires carrying the resummation procedure one order further. At the order displayed here, the nontrivial resonant $A$-cycle monodromy coefficients associated with the $n=1$ resonant solutions are
\begin{equation}
    K^{(1)}_{\pm}=-2(2m_3+1)\mathfrak{t}+\mathcal{O}(\mathfrak{t}^2),
\end{equation}
and contributions proportional to $2m_1+1$ are expected to appear at higher orders in the $\mathfrak{t}$-expansion.

\section{Conclusions}
\label{sec:conclusions}

In this work, we initiated a study of how blow-up equations can be used to investigate special solutions of linear differential equations arising in gauge theory as quantum Seiberg--Witten curves and, more generally, gauge-theoretic solutions of the associated quantum integrable systems. In particular, we focused on orbifold surface-defect partition functions in the $\mathrm{SU}(2)$ gauge theories with $N_f=(n_0,n_1)$ fundamental matter, $n_0,n_1\leq 2$, which provide gauge-theoretic constructions of Floquet-type solutions to the Heun equation and its confluent limits. Such solutions may prove useful in the analysis of periodic spectral problems associated with the Heun equation and its various confluences. 

More specifically, we first derived a new class of blow-up equations \eqref{eq: blowup combined} involving exclusively NS functions, which are satisfied by the gauge-theoretic solutions of the Heun equation and its confluent limits. We then analyzed the analytic structure of the NS functions as functions of the Coulomb-branch parameter $a$ and showed how the resulting blow-up equations can be used to construct resummed NS functions that remain finite at the resonant points $2a=\hbar n$, $n \in \mathbb{Z}$, where solutions that are (anti)periodic with respect to the $A$-cycle monodromy appear.

We then showed how the resonant solutions and their logarithmic companions can be extracted from these resummed NS functions. Beyond providing a gauge-theoretic construction of the resonant solutions themselves, the main advantage of this framework lies in the analytic control it affords in a neighbourhood of the resonant points, which appear as poles in the original NS functions. Finally, we discussed the mass loci on which the resonant $A$-cycle monodromy ceases to be logarithmic and becomes semisimple again, and concluded by illustrating the general construction of resonant solutions and their logarithmic companions in the explicit example of the $N_f=(1,1)$ theory.

This work opens several directions for further investigation:

\begin{itemize}
    \item From a practical standpoint, solving the blow-up equations \eqref{eq: blowup res} is computationally demanding. It would therefore be useful to develop a more efficient method for extracting their solutions. In particular, it would be desirable to identify recursion relations among the resummation functions, although it is presently unclear whether such relations can exist.
    
    \item It would be interesting to determine whether the conjecture \eqref{eq:conj_ad} holds and, if so, to clarify its physical interpretation for the four-dimensional gauge theories, and the two-dimensional effective gauge theories, for example in light of the discussion in \cite{Gorsky:2017ndg}. From a mathematical perspective, the conjecture may also have implications for the geometry of the space of monodromy data, on which $a$ and $a_D$ provide a local system of Darboux coordinates.
    
    \item It would also be interesting to determine, or characterize, the full $\mathfrak{t}$-dependent semisimple locus of the resonant $A$-cycle monodromy beyond the expansion around $\mathfrak{t}=0$ discussed in Section \ref{sec: special mass loci}.
    
    \item Turning to local solutions, the resummation procedure appears to extend to vortex-string partition functions, although this direction requires further investigation. In particular, fixing the regular part of the defect NS prepotentials does not seem to be as straightforward as in the case of orbifold surface-defect partition functions.
    
    \item It would also be interesting to extend the resummation procedure to other gauge theories and investigate its implications for their associated quantum Seiberg--Witten curves and quantum integrable systems. The most immediate four-dimensional generalization is the $\mathcal{N}=2^{*}$ $\mathrm{SU}(2)$ gauge theory and the corresponding Lamé equation, for which the relevant blow-up equations have been proved in \cite{Bershtein:2024kwe}.

    Another natural direction is the extension to five-dimensional $\mathcal{N}=1$ $\mathrm{SU}(2)$ gauge theories, whose quantum Seiberg--Witten curves are finite-difference equations. In this setting, the accessory parameters are encoded in expectation values of BPS Wilson loops, while the corresponding solutions are engineered by codimension-two defect partition functions. Since blow-up equations are available for both classes of observables \cite{Kim:2021gyj, Kim:2025qaf}, it would be interesting to investigate whether they can be used to develop a five-dimensional analogue of the resummation procedure considered here.

    More generally, one may seek to extend the entire construction to four- and five-dimensional gauge theories with higher-rank gauge groups, as well as to quiver gauge theories and the corresponding higher-order quantum curves.

    \item The Heun equation and its confluent limits, together with their gauge-theoretic and CFT avatars, namely $\mathrm{SU}(2)$ NS functions and, through AGT, classical Virasoro conformal blocks, have found a growing range of applications to black-hole quasi-normal modes and other observables \cite{BH1,BH2,BH3,BH4,BH5,BH6,BH7,BH8,BH9}, compact horizonless gravitational objects \cite{CO1,CO2}, holographic correlation functions \cite{HC1}, and holographic geodesic networks \cite{GN1,GN2}. These developments have been facilitated by the gauge-theoretic derivation of the Heun connection formulas \cite{Jeong:2018qpc} and their CFT counterparts \cite{Bonelli:2022ten}. Related resummation procedures have already been used to investigate pole structures and gravitational response functions in some of these settings \cite{RS1,RS2}. It would therefore be interesting to determine whether the resonant solutions and logarithmic companions constructed here admit applications in these and other problems in which the Heun equation plays a prominent role.

    \item Finally, it would be interesting to clarify the implications of the resummation procedure for the associated isomonodromic deformation problem. Resonant values of the $A$-cycle monodromy are loci where the usual coordinates become singular or non-semisimple. The resummed NS functions give regular gauge-theoretic quantities near these loci, while the split eigenfunction separates the resonant solution from its logarithmic companion. This suggests that the resummation may provide a natural regularization of the isomonodromic tau function, or equivalently a preferred local description of the flat connection, near resonant monodromy strata. Along the line of \cite{Jeong:2020uxz}, understanding this relation could shed light on how defect blow-up equations encode the non-semisimple and logarithmic sectors in the corresponding isomonodromic systems.

\end{itemize}

\acknowledgments

We would like to thank M. Bershtein, G. Bonelli, P. Gavrylenko, A. Grassi, D. Guzzetti, C. Iossa, A. Tanzini for helpful discussions and comments. The work of SJ is supported by the Institute for Basic Science under the project IBS-R003-Y3. The work of TP is partly supported by the INFN Iniziativa Specifica GAST, Indam GNFM, the MIUR PRIN Grant 2020KR4KN2 “String Theory as a bridge between Gauge Theories and Quantum Gravity”. Furthermore, TP acknowledges funding from the EU project Caligola (HORIZON-MSCA-2021-SE-01), Project ID: 101086123, and CA21109 – COST Action CaLISTA. Finally, TP gratefully acknowledges the hospitality of CERN and the University of Geneva.

\newpage

\appendix

\section{Nekrasov functions}
\label{app: special functions}

The Nekrasov partition function
$\mathcal{Z}^{(n_0,n_1)}(a,\boldsymbol{m},\varepsilon_1,\varepsilon_2;\mathfrak{q})$ of the $\mathcal{N}=2$ $\mathrm{SU}(2)$ gauge theory with $N_f=(n_0,n_1)$ hypermultiplets, with $n_0,n_1\leq 2$, depends on the Coulomb branch parameter $a$, the $n_0+n_1$ hypermultiplet masses $\boldsymbol{m}$, the two equivariant parameters $\varepsilon_i$, and the instanton counting parameter $\mathfrak{q}$. It factorizes into classical, one-loop, and instanton contributions,
\begin{equation}
\mathcal{Z}^{(n_0,n_1)}(a,\boldsymbol{m},\varepsilon_1,\varepsilon_2;\mathfrak{q}) = \mathcal{Z}_{\text{cl}}(a,\varepsilon_1,\varepsilon_2;\mathfrak{q})\mathcal{Z}^{(n_0,n_1)}_{\text{1-loop}}(a,\boldsymbol{m},\varepsilon_1,\varepsilon_2)\mathcal{Z}^{(n_0,n_1)}_{\text{inst}}(a,\boldsymbol{m},\varepsilon_1,\varepsilon_2;\mathfrak{q}),
\end{equation}
where the classical contribution is
\begin{equation}
\mathcal{Z}_{\text{cl}}(a,\varepsilon_1,\varepsilon_2;\mathfrak{q})=\mathfrak{q}^{-\frac{a^2}{\varepsilon_1 \varepsilon_2}}.
\end{equation}
The one-loop contribution can be conveniently expressed in terms of the Barnes double gamma function\footnote{Throughout this appendix, we mostly follow the conventions of \cite{Jeong:2020uxz}.},
\begin{equation}
\label{eq: double gamma}
\Gamma_2(x;\varepsilon_1,\varepsilon_2)
=\exp\!\left[
-\left.\frac{d}{ds}\right|_{s=0}\,
\frac{1}{\Gamma(s)}
\int_{0}^{\infty} d\beta\,\beta^{s-1}\,
\frac{e^{-\beta x}}{\bigl(1-e^{-\beta\varepsilon_1}\bigr)\bigl(1-e^{-\beta\varepsilon_2}\bigr)}
\right],
\end{equation}
as
\begin{equation}
\mathcal{Z}_{\text{1-loop}}^{(n_0,n_1)}
=\prod_{\pm}\,
\frac{\Gamma_2(0;\varepsilon_1,\varepsilon_2)\,\Gamma_2(\pm 2a;\varepsilon_1,\varepsilon_2)}
{\prod_{i_0=1}^{n_0}\Gamma_2(\pm a-m_{2+i_0};\varepsilon_1,\varepsilon_2)\prod_{i_1=1}^{n_1}\Gamma_2(\pm a-m_{i_1};\varepsilon_1,\varepsilon_2)}\,.
\end{equation}
The instanton contribution is written as a sum over pairs of Young diagrams $\bl=\br{\l^{(0)},\l^{(1)}}$,
\begin{equation}
\mathcal{Z}^{(n_0,n_1)}_{\text{inst}}(a,\boldsymbol{m},\varepsilon_1,\varepsilon_2;\mathfrak{q})= \sum_{\bl}\mathfrak{q}^{|\bl|}\mathcal{E}(-T[\bl]^{(n_0,n_1)}),
\end{equation}
where the full character of the equivariant torus action is
\begin{equation}
T[\bl]^{(n_0,n_1)}=NK^*+q_1 q_2 N^* K-(1-q_1)(1-q_2)KK^*-M^{(n_0,n_1)}K^*,
\end{equation}
with elementary characters
\begin{subequations}
\begin{align}
&N=\sum_{\alpha=0,1}e^{a_\alpha},\\
&K = \sum_{\alpha = 0,1} \sum_{\substack{(i, j) \in \l^{(\alpha)}}} 
    e^{a_{\alpha} + \varepsilon_{1}(i - 1) + \varepsilon_{2}(j - 1)},\\
& M^{(n_0,n_1)} = \sum_{\alpha=1}^{n_0}e^{m_{2+\alpha}}+\sum_{\alpha=1}^{n_1}e^{m_{\alpha}}.
\end{align}
\end{subequations}
To write the one-loop contribution to the orbifold surface defect partition functions considered in this work, we also introduce the single gamma function
\begin{equation}
\label{eq: single gamma}
\Gamma_1(x;\varepsilon_1)
=\exp\!\left[
-\left.\frac{d}{ds}\right|_{s=0}\,
\frac{1}{\Gamma(s)}
\int_{0}^{\infty} d\beta\,\beta^{s-1}\,
\frac{e^{-\beta x}}{1-e^{-\beta\varepsilon_1}}
\right]
=\frac{\sqrt{2\pi\varepsilon_1}}{\varepsilon_1^{\frac{x}{\varepsilon_1}}\,
\Gamma\,\left(\dfrac{x}{\varepsilon_1}\right)}\,.
\end{equation}

The one-loop part of the orbifold surface defect partition function in \eqref{eq: full orbifold} is then
\begin{equation}
    \Psi_{\beta,\,1\text{-loop}}^{(n_0,n_1)}(a,\boldsymbol{m},\varepsilon_1,\varepsilon_2)=\mathcal{Z}^{(n_0,n_1)}_{\text{1-loop}}(a,\boldsymbol{m},\varepsilon_1,\varepsilon_2)\hat{\mathcal{Z}}^{(n_1)}_{\beta,\,\text{1-loop}}(a,\boldsymbol{m},\varepsilon_1),
\end{equation}
where
\begin{equation}
    \hat{\mathcal{Z}}^{(n_1)}_{\beta,\,\text{1-loop}}(a,\boldsymbol{m},\varepsilon_1)=\frac{\prod_{i_1=1}^{n_1}\Gamma_1(a_\beta -m_{i_1};\varepsilon_1)}{\Gamma_1(2 a_\beta ;\varepsilon_1)},
\end{equation}
with $a_0=a$ and $a_1=-a$.

The one-loop bulk and defect Nekrasov functions admit an $\varepsilon_2$-expansion analogous to the instanton contributions in \eqref{eq: NS expansion}. Defining
\begin{equation}
\begin{aligned}
\gamma_2(x;\varepsilon_1) =& \,\varepsilon_1 \left[
\frac{1}{12}+\br{\frac{x}{\varepsilon_1}-1}\log\Gamma\br{\frac{x}{\varepsilon_1}}-\log G\br{\frac{x}{\varepsilon_1}}
\right.\\
&\left.-\log A+\frac{\log\varepsilon_1-1}{2}\,B_2\br{\frac{x}{\varepsilon_1}} \right],
\end{aligned}
\end{equation}
where $A$ is Glaisher's constant, $G$ is the Barnes $G$-function and
\begin{equation}
     B_2(z) = z^2-z+\frac{1}{6},
\end{equation}
is the second Bernoulli polynomial, one finds formally
\begin{equation}
\log\Gamma_2(x;\varepsilon_1,\varepsilon_2) = \frac{1}{\varepsilon_2}\gamma_2(x;\varepsilon_1) +\frac{1}{2}\log \Gamma_1(x;\varepsilon_1) + \mathcal{O}(\varepsilon_2).
\end{equation}
Thus, for the 1-loop part of the twisted effective superpotential we find
\begin{equation}
\begin{aligned}
\mathcal{W}_{0,\, \text{1-loop}}^{(n_0,n_1)}(a,\boldsymbol{m},\varepsilon_1)&=2\gamma_2(0;\varepsilon_1)+\sum_{\pm} \Bigg(\gamma_2(\pm 2a;\varepsilon_1)\\
& \left.-\sum_{i_0=1}^{n_0}\gamma_2(\pm a-m_{2+i_0};\varepsilon_1) -\sum_{i_1=1}^{n_1}\gamma_2(\pm a-m_{i_1};\varepsilon_1)\right),
\end{aligned}
\end{equation}
while, for its first $\varepsilon_2$-correction,
\begin{equation}
\label{eq: W1 1-loop}
    \mathcal{W}_{1,\, \text{1-loop}}^{(n_0,n_1)}(a,\boldsymbol{m},\varepsilon_1)=\frac{1}{2}\log \prod_{\pm}\frac{\prod_{i_0=1}^{n_0}\Gamma\br{\pm \frac{a}{\varepsilon_1}-\frac{m_{2+i_0}}{\varepsilon_1}}\prod_{i_1=1}^{n_1}\Gamma\br{\pm \frac{a}{\varepsilon_1}-\frac{m_{i_1}}{\varepsilon_1}}}{\Gamma\br{\pm \frac{2a}{\varepsilon_1}}},
\end{equation}
Meanwhile, for the defect prepotential, the one-loop part reads:
\begin{equation}
\label{eq: Wb 1-loop}
    \hat{\mathcal{W}}_{\beta,\, \text{1-loop}}^{(n_0,n_1)}(a,\boldsymbol{m},\varepsilon_1)=\mathcal{W}_{1,\, \text{1-loop}}^{(n_0,n_1)}(a,\boldsymbol{m},\varepsilon_1)+\log \frac{\Gamma\br{\frac{2a_\beta}{\varepsilon_1}}}{\prod_{i_1=1}^{n_1}\Gamma\br{\frac{a_\beta}{\varepsilon_1}-\frac{m_{i_1}}{\varepsilon_1}}}.
\end{equation}
In both expressions \eqref{eq: W1 1-loop} and \eqref{eq: Wb 1-loop}, we have omitted constant terms that are irrelevant for our purposes.

To obtain the $\ell$-factors appearing in \eqref{eq: blowup res}, we introduce the $L$-factors
\begin{equation}
    L_{n}^{(n_0,n_1)}= \frac{\mathcal{Z}^{(n_0,n_1)}_{\text{1-loop}}\br{a+n\varepsilon_1,\boldsymbol{m}+\xi_n\varepsilon_1,\varepsilon_1,\varepsilon_2-\varepsilon_1}\mathcal{Z}^{(n_0,n_1)}_{\text{1-loop}}\br{a+n\varepsilon_2,\boldsymbol{m}+\xi_n\varepsilon_2,\varepsilon_1-\varepsilon_2,\varepsilon_2}}{\mathcal{Z}^{(n_0,n_1)}_{\text{1-loop}}(a,\boldsymbol{m},\varepsilon_1,\varepsilon_2)},
\end{equation}
where $\xi_n$ vanishes for $n \in \mathbb{Z}$, and equals $\frac{1}{2}$ for $n \in \mathbb{Z}+\frac{1}{2}$. Using the shift properties of the double gamma function \eqref{eq: double gamma}, these reduce to rational functions of the equivariant parameters,
\begin{equation}
L_{n}^{(n_0,n_1)}(a,\boldsymbol{m},\varepsilon_1,\varepsilon_2) = \frac{\prod_{i_0=1}^{n_0}\prod_\pm \, s_{\pm n-\xi_n}\br{m_{2+i_0}\mp a}\prod_{i_1=1}^{n_1}\prod_\pm \, s_{\pm n-\xi_n}\br{m_{i_1}\mp a}}
{\prod_\pm s_{\pm 2n}(\mp 2a)}
\end{equation}
where:
\begin{equation}
s_n(x)=
\begin{cases}
\prod_{\substack{i,j\ge 0\\ i+j\le n-1}}\br{x-i\,\varepsilon_1-j\,\varepsilon_2},
& n>0,\\[0.6em]
\prod_{\substack{i,j\ge 0\\ i+j\le -n-2}}\br{x+(i+1)\varepsilon_1+(j+1)\varepsilon_2},
& n<-1,\\[0.6em]
1, & n=0\ \text{or}\ n=-1~.
\end{cases}
\end{equation}
The $\ell$-factors are then given by
\begin{equation}
    \ell_{n}^{(n_0,n_1)}(a,\boldsymbol{m}) = L_{n}^{(n_0,n_1)}(a,\boldsymbol{m},1,1) \frac{\hat{\mathcal{Z}}^{(n_1)}_{0,\,\text{1-loop}}(a+n,\boldsymbol{m}+\xi_n,1)}{\hat{\mathcal{Z}}^{(n_1)}_{0,\,\text{1-loop}}(a,\boldsymbol{m},1)}
\end{equation}
which, again using the shift properties of the gamma function in \eqref{eq: single gamma}, can be expressed as rational functions of the equivariant parameters as in equation \eqref{eq: ell factors explicit}.

The regular parts of the resummation ansatz appearing in \eqref{eq: W0 resum}, \eqref{eq: W1 resum} and \eqref{eq: What resum} for the $N_f=(2,2)$ theory can be obtained by studying the large-$a$ behavior of the corresponding instanton prepotentials, and read respectively
\begin{subequations}
\label{eq: regular parts}
\begin{align}
&R_0\left( a,\boldsymbol{m},\hbar;\mathfrak{t}\right)=\frac{a^2}{\hbar} \left[ \frac{\pi K(1-\mathfrak{t}^2)}{K(\mathfrak{t}^2)} + \log \left(\frac{\mathfrak{t}^2}{16}\right) \right] 
+ \log \left(\frac{2 K(\mathfrak{t}^2)}{\pi }\right) \left[\sum_{i=1}^4 m_i\right.\\
&\qquad \qquad \qquad\quad\,\left.+ \frac{1}{\hbar} \sum_{i=1}^4 m_i^2 +\frac{\hbar}{2} \right] 
+ \frac{1}{4\hbar}\log \left(1-\mathfrak{t}^2\right) \left( \sum_{i=1}^4 m_i +\hbar \right)^2,\notag\\
&R_1\left( a,\boldsymbol{m},\hbar;\mathfrak{t}\right)=\frac{1}{\hbar}\br{\sum_{i=1}^4 m_i + \hbar} \left[\log \left(\frac{2 K(\mathfrak{t}^2)}{\pi }\right)+\frac{1}{2}\log \left(1-\mathfrak{t}^2\right)\right],\\
&\hat{R}_0\left( a,\boldsymbol{m},\hbar;z,\mathfrak{t}\right)=\frac{a}{2 \hbar} \left[ \frac{\pi K(1-\mathfrak{t}^2)}{K(\mathfrak{t}^2)} + \log \left(\frac{\mathfrak{t}^2}{16}\right) +2\hat{s}(z,\mathfrak{t})\right]\\
&\qquad \qquad \qquad\qquad+ \frac{2(m_1+m_2)+\hbar}{4 \hbar} \log (1-\mathfrak{t} z)\notag\\
&\qquad \qquad \qquad\qquad+ \frac{2(m_3+m_4)+\hbar}{4 \hbar} \left[ 2 \log \left(\frac{2 K(\mathfrak{t}^2)}{\pi }\right)+\log \left(1-\frac{\mathfrak{t}}{z}\right) \right]\notag,
\end{align}
\end{subequations}
where the non-trivial function $\hat{s}(z,\mathfrak{t})$ can be determined from a large-$a$ WKB analysis of the Heun equation \eqref{eq: Heun (2,2) zt} and is given by
\begin{equation}
    \hat{s}(z,\mathfrak{t})=\int^{z}\left(\frac{1}{\zeta}-\frac{\pi }{2 \zeta K\left(\mathfrak{t}^2\right) \sqrt{(1-\mathfrak{t} \zeta) \left(1-\mathfrak{t}/\zeta\right)}}\right)d\zeta.
\end{equation}
The regular parts for the other cases of $N_f=(n_0,n_1)$ can be obtained from these by taking suitable decoupling limits, as described in Section~\ref{sec: decoupling}. In the previous equations, $K(k^2)$ denotes the complete elliptic integral of the first kind
\begin{equation}
K(k^2)=\int_{0}^{\pi/2}\frac{d\theta}{\sqrt{1-k^2\sin^2\theta}}
=\int_{0}^{1}\frac{dt}{\sqrt{(1-t^2)(1-k^2 t^2)}}\,.
\end{equation}

\section{Solving the \texorpdfstring{$N_f = (1,1)$}{Nf=(1,1)} blow-up equation}
\label{app: example}
In this appendix, we present the first few equations obtained by expanding the blow-up equation \eqref{eq: blowup res}, and solve them explicitly to illustrate the resummation procedure in the $N_f=(1,1)$ case for the resummation ansatz \eqref{eq: What + W1 resum}. At the end, we list the corresponding $h$-functions, obtained through the relation \eqref{eq: relation h-funs}. Throughout this section we set $\hbar = 1$. Before turning to the expansion, we also list the resummation functions $g_{k,j}$ needed in the derivation. Let us define for convenience the function
\begin{equation}
    \gamma_j(\boldsymbol{m},x)=\sqrt{1+4 w_j^{(1,1)}x^2/\zeta_j^2}.
\end{equation}
For notational simplicity, we will suppress one or both of its arguments in lengthy expressions. As already discussed in Section \ref{sec:resummation}, the resummation functions $g_{k,j}$ take the form \cite{Bonelli:2025bmt}
\begin{subequations}
\label{eq:g21_(1,1)}
\begin{align}
& g_{1,j}(\boldsymbol{m},x)=-\frac{1}{x}\left(\log\!\left[\frac{1}{2}+\frac{1}{2}\gamma_j(\boldsymbol{m},x) \right]+1-\gamma_j(\boldsymbol{m},x)\right),\\
& g_{k\geq 2,j}(\boldsymbol{m},x) =\frac{1}{x^{2k-1}}\left[\gamma_j(\boldsymbol{m},x)^{5-2k}Q_{k,j}(\boldsymbol{m},x^2)+P_{k,j}(\boldsymbol{m},x^2) \right]
\end{align}
\end{subequations}
where $Q_{k,j}$ and $P_{k,j}$ are polynomials in $x^2$ of degrees $2k-3$ and $k-1$, respectively:
\begin{equation}
    Q_{k,j}(\boldsymbol{m},x^2)=\sum_{\alpha=0}^{2k-3}q_{k,j,\alpha}(\boldsymbol{m})x^{2\alpha}, \qquad P_{k,j}(\boldsymbol{m},x^2)=\sum_{\alpha=0}^{k-1}p_{k,j,\alpha}(\boldsymbol{m})x^{2\alpha}.
\end{equation}
The coefficients $p_{k,j,\alpha}$ are fixed in terms of the $q_{k,j,\alpha}$ by imposing the condition 
\begin{equation}
    g_{k,j}(\boldsymbol{m},x)= \mathcal{O}(x), \qquad x \to 0,
\end{equation}
while the relevant $q$-coefficients are:
\begin{subequations}
\begin{align}
q_{2,1,0}&=\frac{1}{12} \left(2-\frac{1}{(2 m_1+1)^2}-\frac{1}{(2 m_3+1)^2}\right),\\
q_{2,1,1}&= \frac{3+4\br{m_1+m_1^2+m_3+m_3^2}-16 \br{m_1 m_3 + m_1^2 m_3+ m_1 m_3^2+ m_1^2 m_3^2}}{48 (2 m_1+1) (2 m_3+1)},\\
q_{2,2,0}&=0,\\
q_{2,2,1}&=\frac{2}{9} (2 m_1+1) (2 m_3+1),
\end{align}   
\end{subequations}
\begin{subequations}
\begin{align}
q_{3,1,0}&=\frac{1}{40} \left(-\frac{1}{(2 m_1+1)^4}-\frac{1}{(2 m_3+1)^4}+2\right),\\
q_{3,1,1}&=\frac{1}{80 (2 m_1+1)^3 (2 m_3+1)^3}\left[2 m_1 (m_1+1) (1+2 m_1 +2 m_1^2)\right.\\
&\times \left. (104 m_3 (m_3+1) (1+2 m_3+2 m_3^2)+9)-1 \right] + \frac{9 (2 m_3+1)}{320 (2 m_1+1)^3}, \notag\\
q_{3,1,2} &=\frac{1}{23040 (2 m_1+1)^2 (2 m_3+1)^2} \bigg( 225 +504 \,m_1 + 504 \,m_3 +72\,m_1^2\\
& -11456\,m_1 m_3+72\,m_3^2-864\,m_1^3-56128\,m_1^2 m_3-56128\,m_1 m_3^2-864\,m_3^3 \notag\\
& -432\,m_1^4-89344\,m_1^3 m_3-254144\,m_1^2 m_3^2-89344\,m_1 m_3^3-432\,m_3^4 \notag\\
& -44672\,m_1^4 m_3-396032\,m_1^3 m_3^2-396032\,m_1^2 m_3^3-44672\,m_1 m_3^4 \notag\\
& -198016\,m_1^4 m_3^2-613376\,m_1^3 m_3^3-198016\,m_1^2 m_3^4 -306688\,m_1^4 m_3^3\notag\\
& - 306688\,m_1^3 m_3^4 -153344\,m_1^4 m_3^4 \bigg),\notag\\
q_{3,1,3} &= \frac{1}{11520 (1 + 2 m_1) (1 + 2 m_3)}\bigg( 45+216\,m_1+216\,m_3 +648\,m_1^2\\
&+256\,m_1 m_3+648\,m_3^2 +864\,m_1^3+128\,m_1^2 m_3+128\,m_1 m_3^2+864\,m_3^3\notag\\
&+432\,m_1^4-256\,m_1^3 m_3-2816\,m_1^2 m_3^2-256\,m_1 m_3^3+432\,m_3^4\notag\\
& -128\,m_1^4 m_3-5888\,m_1^3 m_3^2-5888\,m_1^2 m_3^3-128\,m_1 m_3^4-2944\,m_1^4 m_3^2\notag\\
& -11264\,m_1^3 m_3^3-2944\,m_1^2 m_3^4-5632\,m_1^4 m_3^3-5632\,m_1^3 m_3^4-2816\,m_1^4 m_3^4\bigg).\notag
\end{align}
\end{subequations}
We now turn to the equations for the resummation functions $\tilde{h}_{k,j}^{\pm}$ obtained from the blow-up equation \eqref{eq: blowup res}. We will mostly focus on the equations obtained for $\sigma_1=-1$ and $\sigma_1=0$. In particular, for $\sigma_2=1$ and at order $\mathfrak{t}^0$ one finds
\begin{subequations}
\begin{align}
&1 
+ \frac{(1+2m_3)\,x\, \tilde{H}_{1,1}^{-}(\boldsymbol{m},z,-x)}
{ \left(1+\gamma_1(\boldsymbol{m},x)\right) z }
- \frac{1}{2}
\left(1+\gamma_1(\boldsymbol{m},x)\right)
\tilde{H}_{1,1}^{+}(\boldsymbol{m},z,x)
=0,\\
&1 - \tilde{H}_{1,1}^{-}(\boldsymbol{m},z,-x) 
- \left(\frac{1}{2} + m_1\right) x z \, \tilde{H}_{1,1}^{+}(\boldsymbol{m},z,x) = 0,
\end{align}
\end{subequations}
where we define $\tilde{h}_{1,1}^{\pm}=\log \tilde{H}_{1,1}^{\pm}$ in order to linearize the equations. Solving this system we obtain the leading resummation functions
\begin{subequations}
\begin{align}
&\tilde{h}_{1,1}^{+}(\boldsymbol{m},z,x) = \log\,\left(\frac{\gamma_1(\boldsymbol{m},x)+xz(1+2m_1)-1}{xz(1+2m_1)\,\gamma_1(\boldsymbol{m},x)}\right),\\
&\tilde{h}_{1,1}^{-}(\boldsymbol{m},z,x) = \log\,\left(\frac{\gamma_1(\boldsymbol{m},x)+xz(1+2m_1)+1}{2\gamma_1(\boldsymbol{m},x)}\right),
\end{align}
\end{subequations}
which are responsible for the highest-order poles at $2a=\mp1$, appearing at each order in the $\mathfrak{t}$-expansion of \eqref{eq: What + W1 resum}. In the limit $x\to \pm \infty$ one finds
\begin{subequations}
\begin{align}
 \tilde{h}_{1,1}^{+}(\boldsymbol{m},z,x) &=\log\,\left(\frac{1 + 2m_3 \pm \sqrt{(1+2m_1)(1+2m_3)}\, z}{x z (1+2m_1)(1+2m_3)}\right)\\
& \mp\frac{1}{\left(\sqrt{(1+2m_1)(1+2m_3)} \pm z(1 + 2m_1 )\right) x}\;+\;\mathcal{O}\,\left(x^{-2}\right),\notag\\
\tilde{h}_{1,1}^{-}(\boldsymbol{m},z,x) &= \log\,\left(\frac{1}{2}\pm\frac{\sqrt{(1+2m_1)(1+2m_3)}\, z}{2+4m_3}\right)\\
&\pm\frac{1}{\left(\sqrt{(1+2m_1)(1+2m_3)} \pm z (1+ 2m_1)\right) x}\;+\;\mathcal{O}\,\left(x^{-2}\right)\notag,
\end{align}
\end{subequations}
which matches the behavior discussed in Section~\ref{sec: leading order} of the main text. We can then move on to the equations at order $\mathfrak{t}$, which take the form
\begin{subequations}
\begin{align}
&\frac{(1+2m_3)\,x\left(-1 + xz(1+2m_1) - \gamma_1(x)\right)}
{2 z\,\gamma_1(x)\left(1+\gamma_1(x)\right)}\tilde{h}_{2,1}^{-}(z,-x)-\frac{x(r^{(-1)}_0+r^{(-1)}_1 x+r^{(-1)}_2 x^2)}{8\,z^2\,\gamma_1(x)\,(1+\gamma_1(x))^2}\\
&+\frac{(1+\gamma_1(x))\bigl(-1 + xz(1+2m_1) + \gamma_1(x)\bigr)}
{2(1+2m_1)\,x\,z\,\gamma_1(x)}\tilde{h}_{2,1}^{+}(z,x)=0,\notag\\
&4(1+2m_1)\,x\,z\,((1+2m_3) x+z(1+\gamma_1(x)))\tilde{h}_{2,1}^{+}(z,x)+x (r^{(0)}_0+r^{(0)}_1 x)\\
&+\left(8z\,(1+\gamma_1(x))+4(1+2m_1)\,x\,z\,((1+2m_3) x - z(1+\gamma_1(x)))\right)\tilde{h}_{2,1}^{-}(z,-x)=0\notag,
\end{align}
\end{subequations}
where the coefficients $r^{(\sigma_1)}_i$ are given by
\begin{subequations}
\begin{align}
&r^{(-1)}_0=-2\Bigl((1+2m_3)^2 + \bigl(3+2m_1-2m_3-12m_1m_3\bigr) z^2\Bigr)\,(1+\gamma_1(x)),\\
&r^{(-1)}_1=-z(1+2m_3)\Bigl(7+6m_1-2m_3-20m_1m_3+(1+2m_1)^2 z^2\Bigr)\,(1+\gamma_1(x)),\\
&r^{(-1)}_2=-(1+2m_1)(1+2m_3)^2\Bigl(1+2m_3-(1+2m_1)z^2\Bigr),\\
&r^{(0)}_0=-z\Bigl(7+6m_1-2m_3-20m_1m_3+(1+2m_1)^2z^2\\
&\qquad \,\,\,\,-(1+2m_1)(1+2m_3-(1+2m_1)z^2)\,\gamma_1(x)\Bigr),\notag\\
&r^{(0)}_1=(1+2m_1)\Bigl((1+2m_3)^2+(7+m_1(6-20m_3)-2m_3)z^2\Bigr).
\end{align}
\end{subequations}
Solving the equations we find
\begin{equation}
\tilde{h}_{2,1}^{\pm}(\boldsymbol{m},z,x)= \frac{\hat{Q}_{2,1}^{\pm}(\boldsymbol{m},z,x)+\hat{P}_{2,1}^{\pm}(\boldsymbol{m},z,x)\sqrt{1+4 w_1^{(1,1)}x^2}}{\hat{D}_{2,1}^{\pm}(\boldsymbol{m},z,x)},
\end{equation}
where the functions $\hat{Q}_{2,1}^{\pm}$ and $\hat{P}_{2,1}^{\pm}$ are polynomials in $x$ of degree two and one respectively, with coefficients
\begin{subequations}
\begin{align}
\hat{q}_{2,1,0}^+ &=8 (1 + m_1 - 2 m_1 m_3) z^2,\\
\hat{q}_{2,1,1}^+ &=(1 + 2 m_3) (3 +2m_1- 2 m_3 -12m_1 m_3) z\\
& - (1 + 2 m_1) (3 + 2 m_1 - 2 m_3 - 12m_1m_3) z^3,\notag\\
\hat{q}_{2,1,2}^+ &=-(1 + 2 m_1) (1 + 2 m_3)^3 + (1 + 2 m_1)^2 (1 + 2 m_3)^2 z^2,\\
\hat{q}_{2,1,0}^-&=\hat{q}_{2,1,0}^+ z^{-1},\\
\hat{q}_{2,1,1}^-&= -\hat{q}_{2,1,1}^+ z^{-1},\\
\hat{q}_{2,1,2}^-&=(1 + 2 m_1)^2 (1 + 2 m_3)^2 z - (1 + 2 m_1)^3 (1 + 2 m_3) z^3,\\
\hat{p}_{2,1,\alpha}^{\pm}&=-\hat{q}_{2,1,\alpha}^{\pm}, \qquad \alpha=0,1,
\end{align}
\end{subequations}
while the denominators read
\begin{subequations}
\begin{align}
& \hat D_{2,1}^+\br{\boldsymbol{m},z,x}=-4xz (1 + 2 m_1) (1 + 2 m_3)  (2 z + x ((1 + 2 m_3) - (1 + 2 m_1) z^2)),\\
& \hat D_{2,1}^-\br{\boldsymbol{m},z,x}=4x (1 + 2 m_1) (1 + 2 m_3) ((1 + 2 m_3) x - 2 z - (1 + 2 m_1) x z^2).
\end{align}
\end{subequations}
As $x\to \pm\infty$, the two resummation functions above behave as
\begin{subequations}
\label{eq:example_resum_(1,1)}
\begin{align}
\tilde{h}_{2,1}^{+}(\boldsymbol{m},z,x) &\simeq \frac{2 m_3+1}{4 z}\pm\frac{3+2m_1-2m_3 -12m_1m_3}{4 \sqrt{(2 m_1+1) (2 m_3+1)}}\\
& + \left(\frac{3}{4}-\frac{1}{2(1+2m_1)}-\frac{1}{1+2 m_3}- \frac{1}{2}\frac{1}{1\pm \sqrt{\frac{2 m_1+1}{2 m_3+1}}z}\right)x^{-1},\notag\\
\tilde{h}_{2,1}^{-}(\boldsymbol{m},z,x) & \simeq \frac{1}{4} (2 m_1+1) z\pm\frac{3+2m_1 -2m_3-12m_1m_3}{4 \sqrt{(2 m_1+1) (2 m_3+1)}}\\
&+ \left(\frac{1}{4}-\frac{1}{2(1+ 2m_1)}-\frac{1}{1+2 m_3}+ \frac{1}{2}\frac{1}{1\pm  \sqrt{\frac{2 m_1+1}{2 m_3+1}}z}\right)x^{-1},
\end{align}
\end{subequations}
again confirming the expected behavior discussed in Section~\ref{sec: leading order}. Before turning to the order $\mathfrak{t}^2$ equations, it is useful to consider the order $\mathfrak{t}$ expansions of the blow-up equation \eqref{eq: blowup res} for $\sigma_1=-2$ and $\sigma_1=0$. Using the explicit expressions for all resummation functions determined so far, these give
\begin{subequations}
\label{eq: taylor coeff}
\begin{align}
&\partial_x \tilde{h}^+_{1,2}(m_1,m_3,z,0)= -\frac{m_3 (m_3+1)}{2 z^2},\\
&\partial_x \tilde{h}^-_{1,2}(m_1,m_3,z,0)= -\frac{1}{2} m_1 (m_1+1) z^2,
\end{align}
\end{subequations}
which will simplify the $\sigma_1=-1$ and $\sigma_1=0$ equations at the next order. Those order-$\mathfrak{t}^2$ equations then determine the subleading resummation functions $\tilde{h}_{3,1}^{\pm}$ which we do not report here, as their explicit expressions are rather cumbersome. Nevertheless, they are needed in order to determine the next two resummation functions, $\tilde{h}_{1,2}^{\pm}$ and $\tilde{h}_{2,2}^{\pm}$.

Rather than extracting the Taylor coefficients in \eqref{eq: taylor coeff} from additional expansion points, one could proceed differently: solve the order-$\mathfrak{t}^2$ equations for $\tilde h_{3,1}^{\pm}$ while leaving these coefficients as free parameters, and then fix them by imposing the required analytic behavior of the resummation functions (in particular, regularity as $x\to\infty$). This prescription fixes precisely the values in \eqref{eq: taylor coeff}. More generally, similar undetermined Taylor coefficients may appear at intermediate steps and they can be eliminated either by imposing suitable boundary/regularity conditions (e.g.\ vanishing at $x=0$ or regularity at $x\to\infty$), or equivalently by considering further expansion points of the blow-up equation \eqref{eq: blowup res}, as in the previous examples.

Turning to the $\sigma_2=2$ equations, and again considering the expansions for $\sigma_1=-1$ and $\sigma_1=0$, we find at leading order (i.e.\ at order $\mathfrak{t}$)
\begin{subequations}
\begin{align}
&\frac{m_3 (1+m_3) x}{1+\gamma_2(\boldsymbol{m},x)}\tilde{H}_{1,2}^{-}(m_1, m_3, z, -x) + \frac{1 + \gamma_2(\boldsymbol{m},x)}{2} z^2 \tilde{H}_{1,2}^{+}(m_1, m_3, z, x) -z^2=0\\
&2 - 2 \tilde{H}_{1,2}^{-}(m_1, m_3, z, -x) + m_1 (1 + m_1) x z^2 \tilde{H}_{1,2}^{+}(m_1, m_3, z, x)=0
\end{align}
\end{subequations}
where we again perform the substitution $\tilde{h}_{1,j}^{\pm}=\log \tilde{H}_{1,j}^{\pm}$. The system can be solved to give
\begin{subequations}
\begin{align}
&\tilde{h}_{1,2}^{+}(m_1,m_3,z,x)=\log \br{\frac{-\gamma_2(\boldsymbol{m},x)+m_1 (m_1+1) x z^2+1}{m_1 (m_1+1) x z^2 \gamma_2(\boldsymbol{m},x)}},\\
&\tilde{h}_{1,2}^{-}(m_1,m_3,z,x)=\log\br{\frac{\gamma_2(\boldsymbol{m},x)-m_1 (m_1+1) x z^2+1}{2 \gamma_2(\boldsymbol{m},x)}},
\end{align}
\end{subequations}
and these functions account for the highest-order poles at $a=\mp 1$ appearing at every order in the $\mathfrak{t}$-expansion of \eqref{eq: What + W1 resum}. Expanding them for $x\to\pm\infty$, one finds
\begin{subequations}
\begin{align}
& \tilde{h}_{1,2}^{+}(m_1,m_3,z,x)= \log \left(\frac{\pm \sqrt{m_1 (m_1+1) m_3 (m_3+1)}z^2-m_3 (m_3+1)}{m_1 (m_1+1) m_3 (m_3+1) x z^2}\right)\\
& \qquad \qquad + \frac{1}{x}\frac{1}{m_1 (m_1+1) z^2\mp \sqrt{m_1 (m_1+1) m_3 (m_3+1)} } + \mathcal{O}(x^{-2}),\notag\\
& \tilde{h}_{1,2}^{-}(m_1,m_3,z,x)= \log \left(\frac{1}{2}\mp\frac{1}{2}\frac{m_1 (m_1+1) z^2}{\sqrt{m_1 (m_1+1) m_3 (m_3+1)}}\right)\\
& \qquad \qquad - \frac{1}{x}\frac{1}{m_1 (m_1+1) z^2\mp\sqrt{m_1 (m_1+1) m_3 (m_3+1)}}+ \mathcal{O}(x^{-2}).\notag
\end{align}
\end{subequations}
Turning to the next set of equations at order $\mathfrak{t}$, we find:
\begin{subequations}
\begin{align}
&2 m_3 (1 + m_3) x (1 + 2 m_3 - (1 + 2 m_1) z^2) (1 + \gamma_2(\boldsymbol{m},x) + m_1 (1 + m_1) x z^2)\\
&+ 3 m_3 (1 + m_3) x z (1 + \gamma_2(\boldsymbol{m},x) + 
    m_1 (1 + m_1) x z^2) \tilde{h}_{2,2}^{-}(\boldsymbol{m},z,x)\notag \\
&+ 3 z (2 (1 + \gamma_2(\boldsymbol{m},x)) z^2 + m_3 x (-1 - \gamma_2(\boldsymbol{m},x) + 
       m_1 (1 + m_1) x z^2)\notag \\
&+m_3^2 x (-1 - \gamma_2(\boldsymbol{m},x) + 
       m_1 (1 + m_1) x z^2)) \tilde{h}_{2,2}^{+}(\boldsymbol{m},z,x)=0,\notag\\
&(-1 + \gamma_2(\boldsymbol{m},x) -x z^2m_1(1+m_1)) (2 + 4 m_3 - (2  + 4 m_1) z^2 - 
    3 z \tilde{h}_{2,2}^{+}(\boldsymbol{m},z,x))\\
&-3 z (1 + \gamma_2(\boldsymbol{m},x) + 
     x z^2m_1(1+m_1)) \tilde{h}_{2,2}^{-}(\boldsymbol{m},z,x)=0,\notag
\end{align}
\end{subequations}
which can again be solved straightforwardly, giving:
\begin{subequations}
\begin{align}
&\tilde{h}_{2,2}^{+}(\boldsymbol{m},z,x)=\frac{2 \left(1+2 m_3-(1+2 m_1) z^2\right) \left( \left(1-\gamma_2(\boldsymbol{m},x)z^2\right)-m_3 (m_3+1) x\right)}{3 m_1 (m_1+1) x z^5-3 m_3 (m_3+1) x z+6 z^3},\\
&\tilde{h}_{2,2}^{-}(\boldsymbol{m},z,x)=\frac{2 z \left((2 m_1+1) z^2-2 m_3-1\right) \left(\gamma_2(\boldsymbol{m},x)+m_1 (m_1+1) x z^2-1\right)}{3 m_1 (m_1+1) x z^4-3 m_3 (m_3+1) x-6 z^2}.
\end{align}
\end{subequations}
As $x\to\pm\infty$, the two functions behave as
\begin{subequations}
\begin{align}
&\tilde{h}_{2,2}^{+}(\boldsymbol{m},z,x)= \frac{2}{3 z}  \frac{\left(1+2 m_3-(1+2 m_1) z^2\right)}{ m_3 (m_3+1)-m_1 (m_1+1) z^4}\\ 
&\qquad \qquad \qquad \times \left(\pm z^2 \sqrt{m_1 (m_1+1) m_3 (m_3+1)}+m_3(1+m_3)\right)+\mathcal{O}(x^{-1}),\notag\\
&\tilde{h}_{2,2}^{-}(\boldsymbol{m},z,x)=\frac{2z}{3}\frac{\left((2 m_1+1) z^2-1-2 m_3\right)}{ m_1 (m_1+1) z^4-m_3 (m_3+1)}\\
&\qquad \qquad \qquad \times \left(\pm\sqrt{m_1 (m_1+1) m_3 (m_3+1)}+m_1 (m_1+1) z^2\right)+ \mathcal{O}(x^{-1}),
\end{align}
\end{subequations}
and one can also notice that they are related by
\begin{equation}
\tilde{h}_{2,2}^{+}(m_1,m_3,z,x)=\tilde{h}_{2,2}^{-}(m_3,m_1,z^{-1},x).
\end{equation}
Finally, we can determine the corresponding $h$-functions from the relation \eqref{eq: relation h-funs}. As input, we use the functions $\tilde{f}_{1,1}$, $\tilde{f}_{1,2}$, $\tilde{f}_{2,1}$, and $\tilde{g}_{1,1}$, where
\begin{subequations}
\begin{align}
& \tilde{f}_{2,1}(\boldsymbol{m},x)= \frac{3(2m_1+1)^2(2m_3+1)^2-2(2m_1+1)^2-2(2m_3+1)^2}
{32\left(1+(2m_1+1)(2m_3+1)x^2\right)}x^2\!,\\
& \tilde{g}_{1,1}(\boldsymbol{m},x)= -\frac{(1-4m_1m_3)\left(1-\sqrt{1+(2m_1+1)(2m_3+1)x^2}\right)}
{2(2m_1+1)(2m_3+1)x},
\end{align}
\end{subequations}
while the remaining two functions can be found in \eqref{eq: f-tilde functions}. In this way, we recover $h_{1,1}^{\pm}$ and $h_{1,2}^{\pm}$ in the form given in \eqref{eq: resumm leading h}, and
\begin{subequations}
\begin{align}
& h_{2,1}^{\pm}\left(\boldsymbol{m},z,x \right)= \tilde{h}_{2,1}^{\pm}\left(\boldsymbol{m},z,x \right)-\tilde{g}_{1,1}\left(\boldsymbol{m},x \right),\\
& h_{3,1}^{\pm}\left(\boldsymbol{m},z,x \right)= \tilde{h}_{3,1}^{\pm}\left(\boldsymbol{m},z,x \right)-\tilde{f}_{2,1}\left(\boldsymbol{m},x \right),\\
& h_{2,2}^{\pm}\left(\boldsymbol{m},z,x \right)= \tilde{h}_{2,2}^{\pm}\left(\boldsymbol{m},z,x \right).
\end{align}
\end{subequations}


\bibliographystyle{JHEP}

\clearpage
\phantomsection 
\pdfbookmark[1]{References}{References} 

\bibliography{biblio.bib}

\end{document}